\documentclass[twocolumn]{aa}
\usepackage{color}
\usepackage{epsfig}
\usepackage{amsmath}
\usepackage{amssymb}
\usepackage{epic}
\usepackage{float}
\usepackage{array}
\usepackage{multirow}
\usepackage{rotating}
\usepackage{natbib}
\bibpunct{(}{)}{;}{a}{}{,}
\newcommand{\Bvec}{\ovec{B}}
\newcommand{\bvec}{\ovec{b}}
\newcommand{\eBvec}{\ovec{e}_B}

\newcommand{\curl}{\operatorname{curl}}
\newcommand{\dv}{\operatorname{div}}

\renewcommand{\exp}{\operatorname{exp}}

\newcommand{\itover}[2]{\,\hspace{.3mm}#1{\!\hspace{-.3mm}#2}}%verallgemeinern!!

\newcommand{\itdot}[1]{\itover{\dot}{#1}}

\newcommand{\ovec}[1]{{\mbox{\boldmath $#1$}}}
\newcommand{\xvec}{\ovec{x}}
\newcommand{\kvec}{\ovec{k}}

\newcommand{\evec}{\ovec{e}}

\newcommand{\ezvec}{\ovec{e}_z}

\newcommand{\tmax}{{\text{max}}}
\newcommand{\myref}[1]{~\hspace{0pt plus 1pt minus 1pt}\ref{#1}}

\newcommand{\tabref}[1]{Tab.\myref{#1}}

\newcommand{\figsandref}[2]{Figs.\myref{#1} and\myref{#2}}
\newcommand{\figsref}[2]{\figsandref{#1}{#2}}
\newcommand{\figstoref}[2]{Figs.\myref{#1} to\myref{#2}}
\newcommand{\figref}[1]{Fig.\myref{#1}}
\newcommand{\tablref}[1]{Table\myref{#1}}
\newcounter{saveqn}
\newcommand{\alpheqn}{\refstepcounter{equation}\setcounter{saveqn}{\value{equati
on}}%
\setcounter{equation}{0}%
\renewcommand{\theequation}{\mbox{\arabic{chapter}.\arabic{saveqn}\alph{equation
}}}}
 
\newcommand{\reseteqn}{\setcounter{equation}{\value{saveqn}}%
\renewcommand{\theequation}{\arabic{chapter}.\arabic{equation}}}

\def\homfeld(#1,#2)(#3,#4)#5#6%
{\multiput(#1,#2)(#3,#4){#5}{\vector(-#4,#3){#6}}}

\def\quadfeld(#1,#2)(#3,#4)#5%
{\multiputlist(#1,#2)(#3,#4)[#5]{%
{\vector(#4,#3){3.75}},%
{\vector(#4,#3){7}},%
{\vector(#4,#3){9.75}},%
{\vector(#4,#3){12}},%
{\vector(#4,#3){13.75}},%
{\vector(#4,#3){15}},%
{\vector(#4,#3){15.75}},%
{\vector(#4,#3){16}},%
{\vector(#4,#3){15.75}},%
{\vector(#4,#3){15}},%
{\vector(#4,#3){13.75}},%
{\vector(#4,#3){12}},%
{\vector(#4,#3){9.75}},%
{\vector(#4,#3){7}},%
{\vector(#4,#3){3.75}}%
}}

\def\belfelda(#1,#2)(#3,#4)#5%
{\multiputlist(#1,#2)(#3,#4)[#5]{%
{\vector(#4,#3){1.5}},%
{\vector(#4,#3){2.5}},%
{\vector(#4,#3){4}},%
{\vector(#4,#3){7}},%
{\vector(#4,#3){10}},%
{\vector(#4,#3){13.5}},%
{\vector(#4,#3){15.75}},%
{\vector(#4,#3){16}},%
{\vector(#4,#3){15.75}},%
{\vector(#4,#3){13.5}},%
{\vector(#4,#3){10}},%
{\vector(#4,#3){7}},%
{\vector(#4,#3){3}},%
}}

\def\belfeldb(#1,#2)(#3,#4)#5%
{\multiputlist(#1,#2)(#3,#4)[#5]{%
{\vector(#4,#3){2}},%
{\vector(#4,#3){6}},%
{\vector(#4,#3){12}},%
{\vector(#4,#3){15.75}},%
{\vector(#4,#3){16}},%
{\vector(#4,#3){13.5}},%
{\vector(#4,#3){10}},%
{\vector(#4,#3){4}},%
}}

\def\belfeldc(#1,#2)(#3,#4)#5%
{\multiputlist(#1,#2)(#3,#4)[#5]{%
{\vector(#4,#3){2}},%
{\vector(#4,#3){5}},%
{\vector(#4,#3){10}},%
{\vector(#4,#3){14}},%
{\vector(#4,#3){15}},%
{\vector(#4,#3){13.5}},%
{\vector(#4,#3){8}},%
{\vector(#4,#3){3}},%
{\vector(-#4,#3){2}},%
{\vector(-#4,#3){5.5}},%
{\vector(-#4,#3){6}},%
{\vector(-#4,#3){5.5}},%
{\vector(-#4,#3){2}},%
}}

\def\dreibein(#1,#2)#3#4#5#6%	#7%
{%
\put(#1,#2)%
{%
\put(0,0){\vector(\sgn{#3},0){\abs{#3}}}%	x-Achse
\put(0,0){\vector(0,\sgn{#5}){\abs{#5}}}%	z-Achse
\MulFloats{#6}{-2}{\lenone}
\ifnum \sgn{#3} = \sgn{#5}
  \put(0,0){\vector(1,1){#4}}%			y-Achse
  \put( #4, #4){{\small$\,y$}}%   		y-Achsen-Beschriftung
  \put( #6,\lenone){{\tiny$0$}}%	  	Ursprungsbeschriftng	
\else%
  \put(0,0){\vector(-1,-1){#4}}%		y-Achse
  \AddFloats{-#4}{#6}{\lentwo}%
  \put( \lentwo, -#4){{\small$\,y$}}%   	y-Achsen-Beschriftung
  \put( #6,\lenone){{\tiny$0$}}%	  	Ursprungsbeschriftng	
\fi
\put( #3,\lenone){{\small$x$}}%    		x-Achsen-Beschriftung
\put( #6, #5){{\small$z$}}%    			z-Achsen-Beschriftung
}}

\def\zweibein(#1,#2)#3#4%
{\put(#1,#2)%
{\put(0,0){\vector(1,0){#3}}%
\put(0,0){\vector(0,1){#3}}%
\put(0,0){{\circle{#4}}}%
\put(0,0){{\circle{.5}}}%
\put(#3,-#4){{\small$x$}}%
\put(-#4,#3){{\small$y$}}%
\put(-#4,-#4){{\small$z$}}%
}}

\usepackage{calc}
\usepackage{ifthen}

\gdef\NameGdef #1{\expandafter\gdef\csname #1\endcsname}
{\catcode`\p=12
 \catcode`\t=12
 \NameGdef{DropPoints}#1pt{#1}
}
\newlength{\FloatOp}
\newcounter{Exponent}
\def\AddFloats#1#2#3{%
 \setlength{\FloatOp}{#1 pt + (#2 pt)}
 \xdef #3{\expandafter\DropPoints\the\FloatOp}
}
\def\MulFloats#1#2#3{%
 \setlength{\FloatOp}{#1 pt * \real{#2}}
 \xdef #3{\expandafter\DropPoints\the\FloatOp}
}
\def\DivFloats#1#2#3{%
\setlength{\FloatOp}{#1 pt / \real{#2}}
\xdef #3{\expandafter\DropPoints\the\FloatOp}
}
\def\RezFloat#1#2{%
\setlength{\FloatOp}{1 pt / \real{#1}}
\xdef #2{\expandafter\DropPoints\the\FloatOp}
}
\def\SqrFloat#1#2{%
\setlength{\FloatOp}{#1 pt * \real{#1}}
\xdef #2{\expandafter\DropPoints\the\FloatOp}
}
\def\PotFloat#1#2#3{%
\setcounter{Exponent}{#2}
\ifthenelse{\value{Exponent} = 0}{\gdef #3{1.0}}{%
\setlength{\FloatOp}{#1 pt}
\addtocounter{Exponent}{-1}
\whiledo{ \not\(\value{Exponent} = 0\) }{%
\setlength{\FloatOp}{\FloatOp*\real{#1}}
\addtocounter{Exponent}{-1}
}
\xdef #3{\expandafter\DropPoints\the\FloatOp}
}}
\def\gobminus -#1{#1}
\def\abs#1{\ifnum #1 < 0
             \gobminus #1
           \else
             #1
           \fi
          }%
\def\sgn#1{\ifnum #1 < 0
             -1
           \else
              1
           \fi
          }%

\newcommand{\Msun}{\mbox{$M_{\odot}\;$}}
\newcommand{\gcmc}{\text{g}\,\text{cm}^{-3}}

\newcommand{\taugrow}{\tau_{\text{growth}}}
\newcommand{\rhoinit}{\rho_\tin}

\newcommand{\Bzero}{B_{0}}
\newcommand{\Bd}{B_{\text{d}}}
\newcommand{\Bdeins}{\Bd = 10^{13}\,\text{G}}
\newcommand{\Bdzwei}{\Bd = 2 \times 10^{13}\,\text{G}}
\newcommand{\Bdfuenf}{\Bd = 5 \times 10^{13}\,\text{G}}
\newcommand{\lamax}{\lambda^\tmax}
\newcommand{\ombzert}{\omega_{\Bzero}\tau_e}
\newcommand{\ombt}{\omega_B\tau_e}

\newcommand{\imunit}{{\text{i}}}
\newcommand{\tin}{{\text{in}}}
\definecolor{hgrau}{gray}{.75}

\begin{document}

\authorrunning{Rheinhardt, Konenkov \& Geppert}
\titlerunning{Hall--Instability in Neutron stars}

% \thesaurus{08.14.1;   % Stars: neutron
%              08.16.6;   % (Stars:) pulsars: general
%              02.13.1    % Magnetic fields
%             }

\title{The Occurrence of the Hall--Instability in  
Crusts of Isolated Neutron Stars}

\author{M. Rheinhardt\inst{1}, D. Konenkov\inst{2} \and U. Geppert\inst{1} }
\institute{Astrophysikalisches Institut Potsdam,
           An der Sternwarte 16, 14482 Potsdam, Germany;
           \email{urme@aip.de} 
       \and
           Ioffe Physico-Technical Institute, 194021 Af. F., Politechnicheskaya 26,
           St. Petersburg, Russia\\
           \email{dyk@astro.ioffe.rssi.ru}
           }

\date{Received date ; accepted date}

\abstract{
In former papers we showed that during the decay of a neutron star's magnetic
field under the influence of the Hall--drift, an unstable rise of small--scale field
structures at the expense of the large--scale background field may happen.
This linear stability analysis was based on
the assumption of a uniform density throughout the neutron star crust, whereas in reality
the density and all transport coefficients vary by many orders of magnitude. 
Here, we extend the investigation of the Hall--drift induced instability
by considering realistic profiles of density and chemical composition,
as well as background fields with more justified radial profiles.
Two neutron star models are considered differing primarily in the assumption on the core matter equation
of state. For their cooling history and radial profiles of density and composition
we use known results to infer the conductivity profiles. These were fed into  
linear calculations of a dipolar field decay starting from various initial configurations.
At different stages of the decay, snapshots of the magnetic fields at the equator were taken to yield
background field profiles for the stability analysis.
The main result is
that the Hall instability may really occur in neutron star crusts.
Characteristic growth times are in the order of $\lesssim 10^4\ldots10^6$ yrs depending
on cooling age and background field strength.
The influence of the equation 
of state and of the initial field configuration is discussed. 
}

\maketitle

\section{Introduction}

Neutron stars (NSs) are carriers of the strongest magnetic fields which
occur in nature.
But, astonishingly enough, most of the quantitative studies of magnetic field decay
in NS crusts consider only the linear induction equation, i.e., a 
field decay caused solely by Ohmic dissipation \citep[see, e.g.,][]{UCS94,UK97,PGZ00}.

\citet{YS91} showed
that in a two-component plasma the resistivity components
parallel and perpendicular to the magnetic field coincide, as a result of which in turn
the ambipolar drift disappears. This result is applicable to the electrons in the
fully ionized crystal matter of the NS crust, too \citep[see][]{US99}.
Further, convective motions will not exist since the crust is almost completely
crystallized after, say, $10^4$ yrs. 
Therefore, the magnetic field evolution in the crust is solely 
determined by Ohmic diffusion and the so--called Hall--drift 
where the latter is the only non--linearity in this process
%which occurs in a crystallized and magnetized NS crust is Hall--drift
(if the weak and therefore weakly nonlinear back--reaction of the magnetic field on the
conductivity tensor via Joule heating is discounted).
Recently, two of us \citep{RG02,GR02} showed that for a
(large--scale) magnetic background field characterized by a sufficiently curved radial profile
beyond a certain marginal field strength, the Hall--drift may cause 
an unstable growth of small--scale perturbations.
For a homogeneous medium these 
are concentrated towards the boundary adjacent to vacuum.

We performed this stability analysis based on the linearized induction
equation with Hall--drift for a homogeneous plane slab of finite thickness, 
infinitely extended in both horizontal dimensions and bound by vacuum and a medium of infinite
conductivity at its upper and lower sides, respectively. 
While this model is perhaps an acceptable approximation of the 
NS crust with respect to geometry, the assumption of a spatially uniform density and chemical composition, 
which result in a uniform scalar conductivity and Hall--drift coefficient, 
is surely a very crude one.
Actually, the density and thus the coefficients 
depending on it may vary throughout the crust by many orders of magnitude
\citep[see, e.g.,][]{PGZ00}.
Less pronouncedly, the chemical composition (that is, mass number $A$ and atomic number $Z$) varies, too. 
Additionally, the scalar conductivity and the Hall--drift coefficient %magnetization parameter
are dependent on the temperature and the impurity concentration.

The effect of the Hall--drift on the magnetic field evolution in NSs
has been considered by a number of authors \citep{GR92,VCO00,US95,US99,SU95,SU97,NK94,HUY90,HR02}.
The only study, however,
which includes quantitatively the effect of the crustal density stratification
in this context is the one by \citet{VCO00}.
They showed that due to a density gradient, the Hall--drift may create current sheets, 
where the field can be dissipated very efficiently. However, the occurrence of
the mentioned Hall--instability could not be detected by them because in their
linear analysis they considered a uniform
background field only, i.e., a field which will not become unstable
regardless which strength it has. In considering the non--linear evolution of the toroidal
field they neglected the coupling with the poloidal one, thus making an instability
impossible either, see \eqref{indeqst} below.

Observational evidence for a decay of the large--scale (dipolar) magnetic field of middle
aged pulsars, being drastically accelerated in comparison with the purely Ohmic decay, has been
discussed in \citet{GR02}. Our reasoning there is based on the detection of braking indices 
greater than three by \citet{JG99}, who used measurements of the rotational period $P$, and its
temporal derivative $\dot{P}$, dating from different observational epochs. As a typical result
we found that the decay times inferred from $P$ and $\dot{P}$ may be smaller than $10^{-4}$ times
the (estimated) Ohmic decay time. 
A possible explanation for such a rapid field decay may be the Hall--instability which drains
magnetic energy out of the dipolar field and uses it for the build up of
small--scale magnetic field structures, which eventually decay by Ohmic dissipation.

The occurrence of small--scale magnetic field structures in the crustal layers of NSs may cause
further effects which are potentially observable. Beyond the decrease of the braking power of 
the magnetic field, the increased small--scale Lorentz forces can trigger a cracking of the crystallized
crust. Also, due to a rapid and spatially concentrated Ohmic decay of the small--scale field 
structures, hot spots may appear in the surface layers.
Recent observations, both in the X--ray \citep{BSP03}, and in the radio range \citep{GMM02}, support
the idea of the existence of strong small--scale magnetic field structures at the NS surface.

Of course, none of the mentioned phenomena
can be unambiguously attributed to
the effect of the Hall--instability. Some possible alternative explanations are discussed
in \citet{GR02}.

However, the simultaneous existence of a sufficient curvature of the background field profile
{\em and} of a sufficiently large magnetization parameter $\ombt$ related to that field
(where $\ombt>1$ is a necessary condition)
will very likely lead to the appearance of the Hall--instability.
How vigorously it
develops for a realistically modelled crust and at which strength it saturates, remain as open
questions. Also unanswered is,
how
far it may be hidden by other effects of the field evolution as, e.g., the ``normal" Hall
cascade.
The first of these questions  will be addressed here.

One of the most convincing indicators for the importance of the Hall--drift in 
the crust is the evolution of the magnetization parameter $\ombt$ there. 
In Fig. 1 of \citet{GR02} this quantity is shown for different temperatures, that is, for different ages
as a function of density. These results are based on
linear field decay calculations for a standard NS model with a medium EOS
as presented by \citet{PGZ00}. 
Clearly, as soon as the magnetic field strength exceeds $10^{12}$~G, $\omega_B\tau > 1$ 
in some regions of the crust, and the Hall--drift  begins to dominate
the Ohmic decay.

In order to answer the question whether at all and how intensive the
Hall--instability occurs in real NS crusts, we will consider crustal density
profiles resulting from NS models based on stiff and medium equations of 
state (EOSs) of the core matter.
As background magnetic fields which must -- in comparison with the unstable field
perturbations -- evolve very slowly, we use those,
calculated by the methods described in \citet{PGZ00}. The
evolution of those large--scale fields is affected only by the density profile of the
NS crust
and its cooling history, both determined essentially by the EOS.
Furthermore, the crustal field evolution is affected by the chemical composition
and impurity concentration within the crust, as well as by the initial
strength and structure of the field which, in turn, reflect the 
processes at birth of the NS. These results are
insofar incomplete, as they just do {\em not} take into account the very efficient drain of magnetic
energy from the background field during the period of the Hall--instability.
They should, however, provide hints under which conditions, more realistic
than those considered up to now, an episode of strongly non--linear magnetic
field decay may take place.

This paper follows the lines of thought as presented by 
\citet{RG02}, and is organized as follows: 
In the next section the basic equations of
the model are derived, and the assumed properties of the crustal matter are described  
together with the background field profiles we used.
Section \ref{results} briefly outlines the method of solving the eigenvalue problem and provides
and discusses the numerical results.
Conclusions are drawn in Sect. \ref{conclus} with a focus on possible
observational consequences of the Hall--instability.

\section{Description of the model}
\label{model}
\subsection{Basic equations, geometry and boundary conditions}
\label{baseq}
In the absence of convective motions and of ambipolar diffusion, conditions 
found in the solid crust of NSs, the equations governing the
magnetic field evolution read
\begin{equation}
\hspace*{-.3cm}\begin{aligned}
\phantom{.}&\,\itdot{\vec B} = -  c~~\mbox{curl}\,\bigg(\frac{c}{4\pi\sigma_0}
\big(\mbox{curl}\,{\vec B}+ \omega_B{\tau_e}(\mbox{curl}\,{\vec B}
\times{\eBvec})\big)\bigg)\\
\phantom{.}&\dv {\vec B} = 0\; ,
\end{aligned}
\label{indeq}
\end{equation}
where $\tau_e$ is the relaxation time of the electrons and 
$\omega_B= e|\vec B|/m_e^* c$ the electron Larmor frequency, with $e$ being
the elementary charge,
$m_e^*$ the effective mass of an electron, and $c$ the speed of light. $\eBvec
$ is the unit vector in $\vec B$--direction. The scalar conductivity $\sigma_0$ is given by
\begin{equation}
 \sigma_0 = \frac{Z}{A} \frac{e^2\rho(1-X_{\text{n}})}{m_u m_e^*}\tau_e\;.
 \label{sig0}
\end{equation}
Here $\rho$ denotes the (depth depending) density of the crustal matter, and $m_u$ the atomic mass unit. 
By virtue of their direct dependences on density, both $m_e^*$ and $\tau_e$ depend on the crustal depth, too.
The mass and atomic numbers $A$ and $Z$, respectively, also depend
on depth as well as the fraction of free neutrons $X_{\text{n}}$ \citep[see, e.g.,][]{NV73,HZ90}.
Eq. \eqref{indeq} lets one expect that the
Hall--drift described by the term $\omega_B{\tau_e}(\mbox{curl}\,{\vec B}\times{\eBvec})$ may become
important for the magnetic field decay only if 
$\omega_B{\tau_e} \gtrsim 1$.
%The further mathematical treatment is as follows:

Linearization of \eqref{indeq} with respect to a background field (= reference state)
$\Bvec_0$ yields:
\begin{equation}
\begin{aligned}
\phantom{.}&\begin{aligned}\dot{\bvec} = &-\curl\,(\eta\,\curl\bvec) \\
                              &-\curl\big(\alpha\,(\,\curl\Bvec_0
                              \times\bvec+\curl\bvec\times\Bvec_0\,)\big)
 \end{aligned}\\
\phantom{.}&\dv \bvec = 0 
\end{aligned}\label{indeqlin}
\end{equation}
describing the behavior of small perturbations $\bvec$ of the reference
state. Here, the magnetic diffusivity 
$\eta$, and the Hall coefficient $\alpha$ are given by 
\begin{equation}
\eta=\frac{c^2}{4\pi\sigma_0}\;,\quad
\alpha=\eta\frac{\ombt}{|\vec B|}=\frac{A}{Z}\frac{m_u c}{4\pi e\rho(1-X_{\text{n}})}\;.
\label{rescoeff}
\end{equation}
Note, that the Hall parameter $\alpha$ is time--independent for a non--accreting
NS while $\eta$ depends strongly on the crustal temperature via $\tau_e$ and, therefore, on time.
In general, $\Bvec_0$ is a {\em decaying} field, i.e., time--dependent, too.
For the instability analysis we nevertheless want to treat them as constants.
Then the following necessary condition has to be satisfied to take the results for valid: The growth time of an unstable mode
has to be significantly shorter than all the characteristic times of changes of coefficients in \eqref{indeqlin}
especially significantly shorter than the decay time of the background field.
In the following this restriction will be referred to as the
``background dynamics permissibility condition".
 
Employing the arguments from dynamo theory quoted in \citet{RG02}, one can
conclude that $\alpha\curl\Bvec_0$ must not be a homogeneous field. More precisely,
when interpreting this term as a velocity field, it must not be a rigid body motion, 
in order to be capable of delivering energy to $\bvec$.
Hence, in contrast to the homogeneous density model, now {\em any inhomogeneous} $\Bvec_0$ may 
potentially enable the instability. 

Let us now specify the geometry of our model. Idealizing the spherical shell of the NS crust,
we consider a plane slab which is infinitely extended
both into the $x$-- and $y$--directions, but has a finite thickness $d$ in
$z$--direction. We identify $z$ with the crustal depth being zero at the surface of the NS.
The background field is assumed to be
parallel to the surfaces of the slab pointing, say, in $x$--direction and to be 
dependent on the depth $z$ only, i.e., 
\begin{equation}
\Bvec_0 = B_0 \evec_x = f(z) \evec_x\;,\label{b0}
\end{equation} 
where in comparison to the homogeneous density model the minimum demand on $f(z)$
can now be relaxed from quadratic to linear in $z$, if only $\alpha$ is not a constant (see above).
Note, that due to this assumption $\alpha\curl\Bvec_0\times\Bvec_0$
represents a gradient. Thus the unperturbed evolution of the
background field is not at all affected by the Hall--drift; in the
absence of any electromotive force it would decay purely ohmically.

XXX The choice \eqref{b0} is not only motivated by the sake of simplicity, but to some
extent justified by the physical conditions during the proto--NS stage (after the end of a possible
field--generating phase): As long as
the layers which later on form the crust are liquid, the magnetic field adjusts to
be close to a magnetostatic equilibrium. Then it approximately obeys the condition
\begin{equation}
\curl \left(\frac{1}{n_e} \curl\Bvec \times\Bvec\right) = \vec{0}
\label{forfree}
\end{equation}
with $n_e$ being the electron number density.
\citep[see][ Sect. 14.1]{TD93}\footnote{There exist even dipolar fields in spherical shells satisfying \eqref{forfree} with $n_e = n_e(r)$ exactly.}.
As the Hall--coefficient $\alpha$
can be written in the form $\alpha=c/4\pi n_e e$ \citep[see, e.g.,][]{GR92} this is at the same time the condition for the Hall e.m.f.
in \eqref{indeq} to be a gradient and thus ineffective. However, as the modes of free decay in the
crystallized crust are in general violating \eqref{forfree}, the crustal field will tend to 
deviate increasingly from the magnetostatic configuration. But at least for early stages,
one may suppose that although $\ombt$ might already be bigger than unity, the effect of the
Hall term on the background field is dominated by ohmic dissipation. XXX

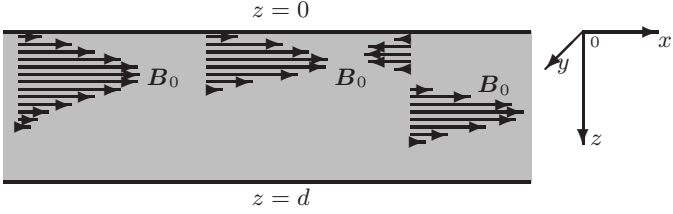
\begin{figure}
\setlength{\unitlength}{1.mm}
\hspace*{-1cm}\begin{picture}(90,25)
\thicklines
\put(10,1){\colorbox{hgrau}{\makebox(68,18){}}}
\put(43,-3){{\small $z=d$}}
\put(10,0){\line(1,0){70}}
\put(10,20){\line(1,0){70}}
\put(43,22){{\small $z=0$}}
\belfelda(12,9)(0,1){l}\put(29,13){$\Bvec_0$}
\belfeldb(37,14)(0,1){l}\put(54,13){$\Bvec_0$}
\belfeldc(64,7)(0,1){l}\put(73,12){$\Bvec_0$}
\dreibein(87,20){10}{5}{-15}{1}
\end{picture}

\vspace{2mm}
\caption{\label{geometrie} Sketch of the model geometry ($\Bvec_0$ - background fields).
In $z>d$ a perfect conductor, in $z<0$ vacuum is assumed.}
\end{figure}

In \figref{geometrie}, the model geometry is shown with three different depth profiles
of $\Bvec_0$, depicting qualitatively some of those we employed.
They will be specified later.

We decompose a perturbation $\bvec$ into
poloidal and toroidal constituents, $\bvec = \bvec_p+\bvec_t$, which can
be  represented by scalar
functions $S$ and $T$, respectively, by virtue of the definitions
\begin{equation}
\hspace*{-7mm}\bvec_p = -\curl\,(\,\ezvec \times \nabla S)\,,\quad 
\bvec_t = -\ezvec \times \nabla T\,, 
\hspace*{-1cm}\label{poltor}
\end{equation}
ensuring $\dv\bvec=0$ for arbitrary $S,T$.

For the sake of simplicity, we will confine ourselves to the
study of plane wave solutions with respect to the $x$-- and $y$--directions,
thus making the ansatz
\begin{equation}
\left\{\begin{aligned}
S\\
T
\end{aligned}\,\right\}(\xvec,\tau) = 
\left\{\begin{aligned}
s\\
t
\end{aligned}\,\right\} (z) \exp{(i\tilde{\kvec}\tilde{\xvec} + p\tau)}\,,
\label{transf}
\end{equation}
where $\tau$ denotes the time, $\tilde{\kvec} = (k_x,  k_y)$, $\tilde{\xvec} = (x,y)$
and $p$ is a complex time increment.

Inserting \eqref{poltor} with \eqref{transf} into \eqref{indeqlin}, we finally obtain two coupled ordinary differential equations for the
scalars $s$ and $t$:
\begin{equation}
\hspace*{-15mm}\begin{aligned}
pt &- \eta(t''-\tilde k^2t) -\eta't' &\hspace*{-2mm}+ \imunit\,\alpha &k_x(f''s+\tilde k^2fs-fs'')\\
&&+\, \imunit\alpha'&(k_yft+k_xf's)=0\\
ps &- \eta(s''-\tilde k^2s) &\hspace*{-2mm}+ \imunit\,\alpha&(k_xft-k_yf's)=0\;,
\end{aligned}
\hspace{-10mm}\label{indeqst}
\end{equation}
where the dash denotes the derivative with respect to 
$z$, and $\tilde k^2 = k_x^2+k_y^2$.
In comparison with the corresponding Eqs. (6) in \citet{RG02}, the terms $\propto$ $\eta'$ and $\alpha'$
occur additionally, but they don't affect the energetic conclusions relevant for the existence
of the instability.

When completed with appropriate boundary conditions, these equations define an
eigenvalue problem with respect to the time increment $p$.
The boundary conditions chosen here are transition to vacuum at
$z=0$, and to a perfect conductor at $z=d$, respectively, mimicking the superconducting core
below the bottom of the crust and an atmosphere with low conductivity above its surface.
They read
\begin{equation}
\hspace*{-.1cm}\begin{aligned}
s' +\tilde k s&=t&=0 &\quad\text{for}\quad z=0\\
s  &= \eta\,t' - \imunit k_y \alpha f\, t\, &=0 &\quad\text{for}\quad z=d \;.
\end{aligned}\hspace{-1cm}
\label{BCst}
\end{equation}
These conditions are equivalent to the requirements that all components of the magnetic field
are continuous across the vacuum boundary, and that neither the normal
component of the magnetic field nor the tangential components of the electric field
penetrate the core. For details see \citet{GR02}.

The signs of the wavenumbers are irrelevant for the eigenvalues of \eqref{indeqst} with \eqref{BCst},
since the transformations
\begin{equation}
\begin{aligned}
k_x\rightarrow -k_x\,,\quad p\rightarrow p  \,,\quad &s\rightarrow s   \,,\quad &t\rightarrow -t\\
%\intertext{and}
\hspace*{-5cm}\text{and}\\
k_y\rightarrow -k_y\,,\quad p\rightarrow p^*\,,\quad &s\rightarrow s^* \,,\quad &t\rightarrow -t^*
\end{aligned}\hspace{-2cm}\label{invar}
\end{equation}
hold ($*$ is the complex conjugate).
Thus it is sufficient to consider the quadrant $k_x,k_y > 0$ of the $k_x$--$k_y$--plane only. On the basis
of \eqref{invar}, it can be easily concluded that changing the sign of $f$ is irrelevant, too, because
\begin{equation}
f\rightarrow -f\,,\quad p\rightarrow p^*\,,\quad s\rightarrow s^* \,,\quad t\rightarrow t^*\; .
\end{equation}
 
As can be inferred from \eqref{indeqst} using standard energy arguments all solutions
for $k_x=0$ are damped oscillations, being either purely toroidal or purely poloidal. However,
for $k_y=0$
growing solutions with their poloidal and toroidal parts mutually coupled seem well possible
if $\Im\{k_x\int_0^d  (\alpha f/\eta)\, (s^* t)\, dz \} > 0$.

\subsection{Crustal matter properties}
The main goal of the present paper is to enhance our knowledge on the conditions under which
the Hall--instability occurs. As one step in this direction we consider a stratified crust characterized by a 
{\em realistic} density profile instead of a homogeneous crust.
\begin{table}[h]
\begin{center}
\begin{tabular}{@{}lcc@{}}
                   \hline
                   \hline\\*[-2mm]
                            & FP	        & PS		    \\*[1mm]
                   \hline\\*[-1.5mm]
EOS                         & medium	        & stiff 	    \\
compactness $GM/c^2R$ 	    &  0.2	        & 0.13  	    \\
central density ($\gcmc$)   &$1.17\cdot 10^{15}$&$3.64\cdot 10^{14}$\\
NS radius       (km)        &  10.64	           &16.4	    \\
crust thickness (m)         & 730	        & 3860  	    \\*[1mm]
                            &  Friedman \&      & Pandharipande     \\
reference		    &  Pandharipande,   & \& Smith,	    \\
			    &	1981 		&   1975	    \\*[1mm]
                   \hline
\end{tabular}
\end{center}
\caption{\label{models}Properties of the Pandharipande--Smith (PS) and Friedman--Pandharipande (FP) NS models; $M_{NS}=1.4 \Msun$.}

\vspace{-5mm}
\end{table}

The resistivity
tensor is strongly dependent on density, temperature, and chemical composition of the crust 
(see Eqs. \eqref{sig0}, \eqref{rescoeff}; $\tau_e$, as far as collisions with phonons are concerned depends on temperature, too, see \citet{UY80}).
In turn, these quantities are in their temporal behavior
mainly determined by the EOS of the NS matter, especially that of the core.
Therefore, one has first
to make a choice among the 
multitude of proposed core EOSs \citep[see, e.g.,][]{vR91}.
We decided to consider two of them, which are generally
accepted to be typical representatives of a medium \citep[][ henceforth FP]{FP81}
and a stiff \citep[][ henceforth PS]{PS75} EOS,
respectively, thus probably covering an essential part of the observed NSs.
\citep[The extreme soft EOSs, e.g., that derived by \citet{BPS71}, are now considered unlikely to be realized in nature, see][]{PGZ00}
Further on, we specify the crust in the FP case to consist of cold catalyzed 
matter, and in the PS case to be dominated by matter accreted and processed in the past. The corresponding EOSs of the crustal matter and its chemical compositions were derived in \citet{NV73} and \citet{HZ90}, respectively.
Since these two models are referred to and utilized in numerous papers, we thus hope to facilitate   
discussion and application of our results.
\begin{figure}[h]
\begin{center}
\epsfig{file=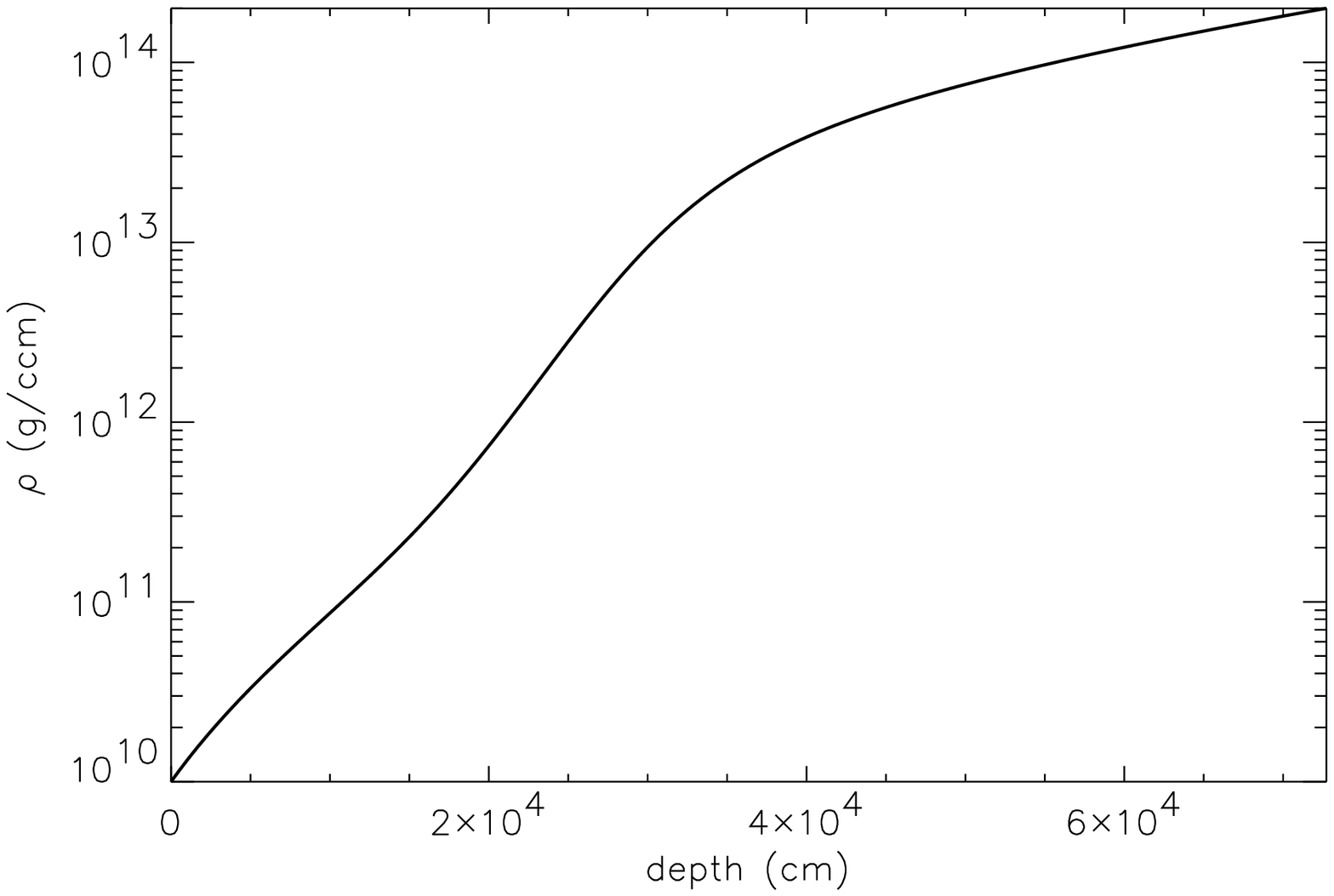, width=.9\linewidth}\\
\hspace*{4mm}\epsfig{file=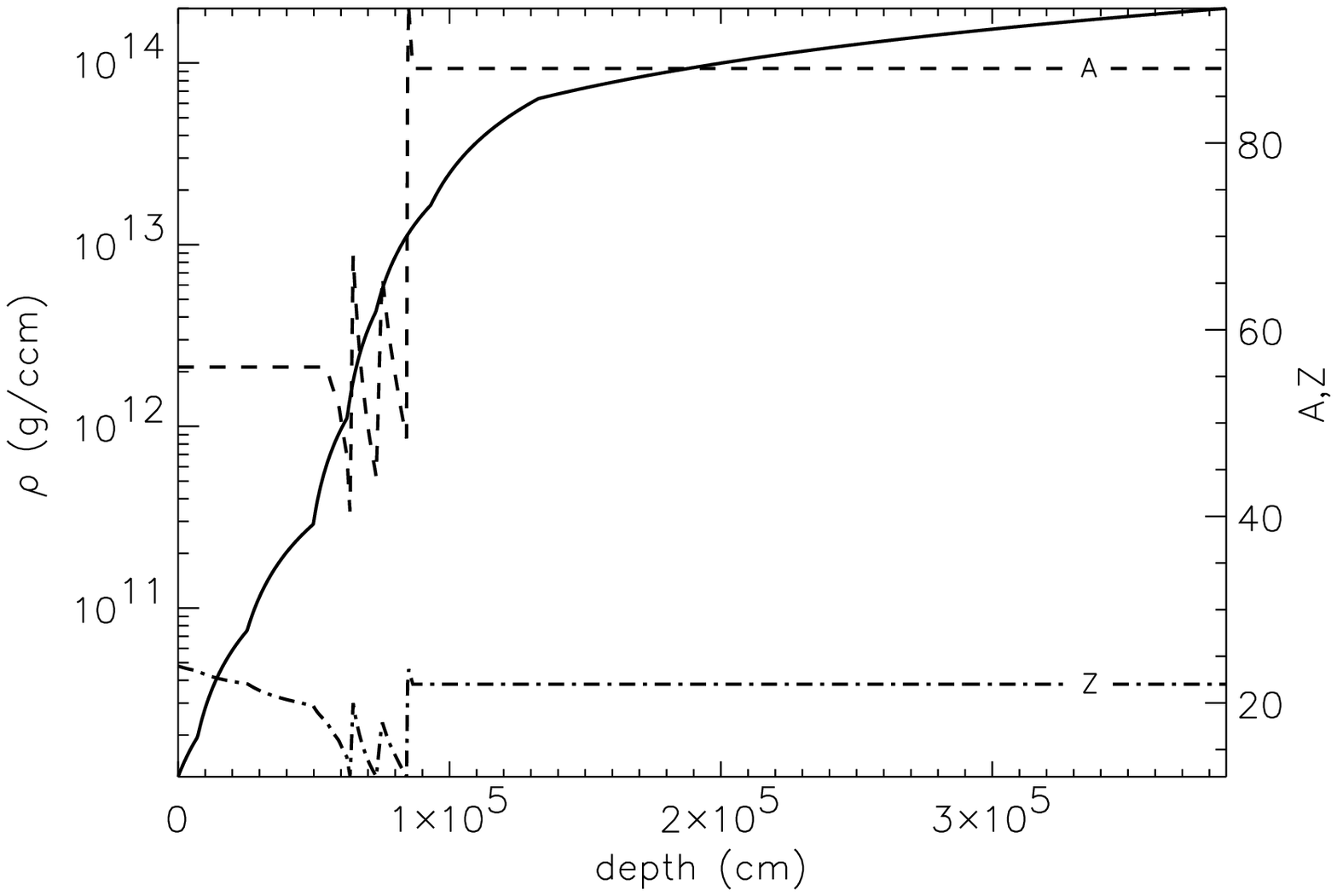, width=\linewidth}
\end{center}
\caption{\label{rhoprofs}Radial density (solid) profiles for the FP (top) and PS (bottom) models.
Additionally, for the latter the chemical composition, $A$, $Z$, (dashed and dash--dotted, respectively) is given.}
\end{figure}

Some characteristic properties of the selected models
resulting from their density structure specified for $M_{NS}=1.4 \Msun$, are summarized in Table \ref{models};
Figure \ref{rhoprofs} shows the density profiles.
Additionally, the profiles of the chemical composition, that is, $A$ and $Z$,
for the PS model as derived from the data in \citet{HZ90} are given.
There exist a number of discrete depths (i.e., densities and pressures)
where the preferred species of nuclei change almost abruptly
causing discontinuities in
$A$ and $Z$.
For the FP model, the corresponding data were given only implicitly in the form of a smooth fitting 
function for $\rho(z)$.
In both cases the crust--core boundary was assumed to be at a density of $2\times10^{14} \gcmc$ whereas the 
crust surface was defined by the density $10^{10} \gcmc$. 

\begin{figure}
\epsfig{file=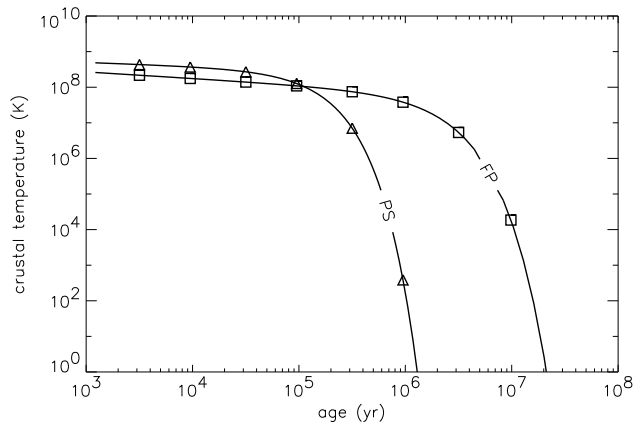, width=\linewidth}
\caption{\label{coolcurvs}Cooling curves for the FP (squares) and PS (triangles) models.
The symbols denote the ages at which the stability analysis is performed.}
\end{figure}
The most important of 
the EOS's direct consequences is the degree of compactness:
The smaller compactness (i.e., larger radius and crust thickness) of the PS model
compared with the FP model of equal mass, results from its stiffer EOS.
Correspondingly, the latter model cools down significantly
slower than the former for ages higher than $10^5$ yrs, since a large compactness (strong gravity)
inhibits the heat transfer \citep[see, e.g.,][]{MTW73}.
\figref{coolcurvs} shows the cooling curves for both models as calculated by \citet{vR91}. 
Throughout this paper we confine ourselves to NSs old enough
(say, some $10^3$ yrs) that the crustal temperature can be considered uniform, and the density profile
is no longer changing in time.  

\begin{figure}[t]
\epsfig{file=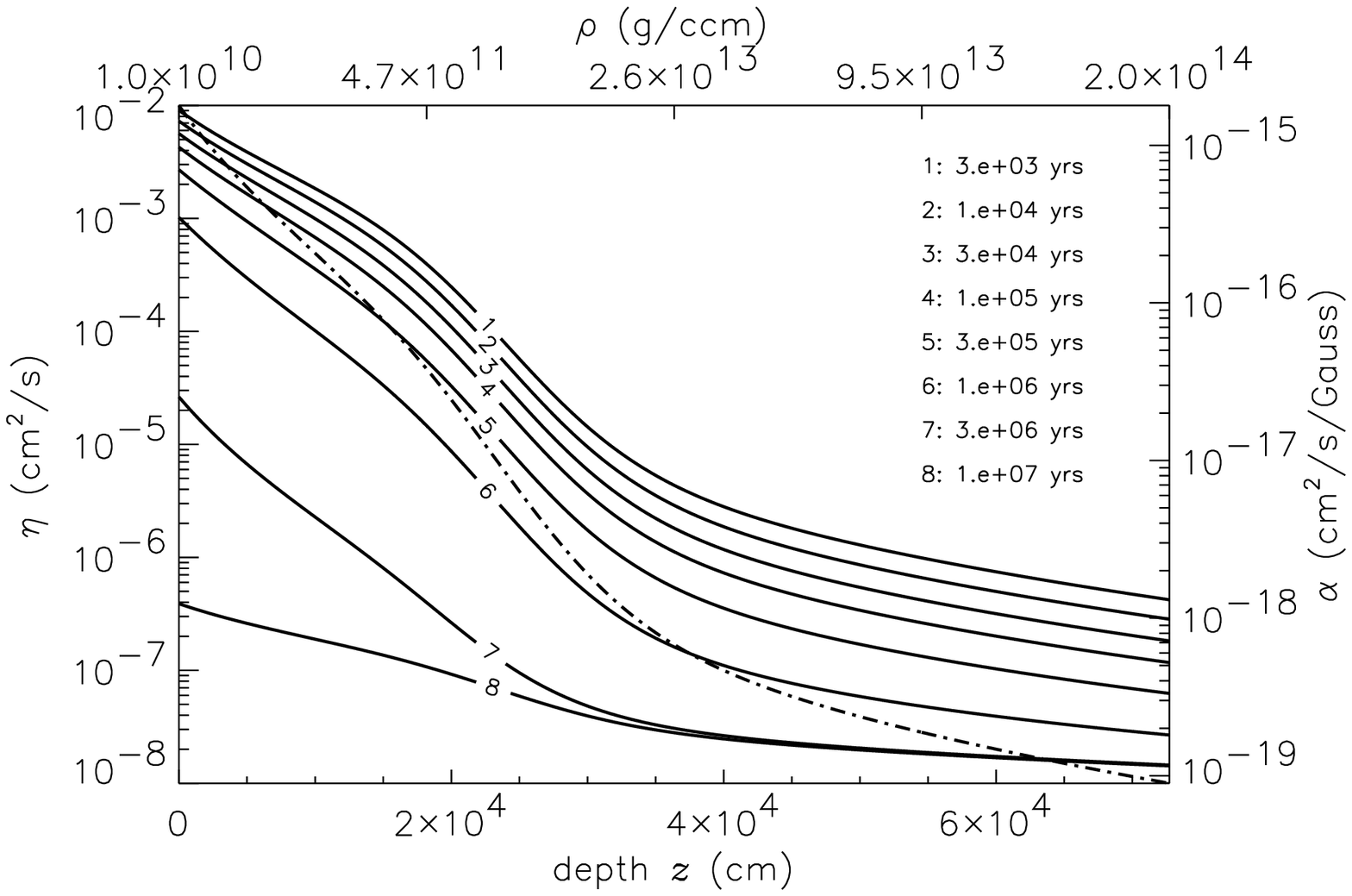, width=\linewidth}
\epsfig{file=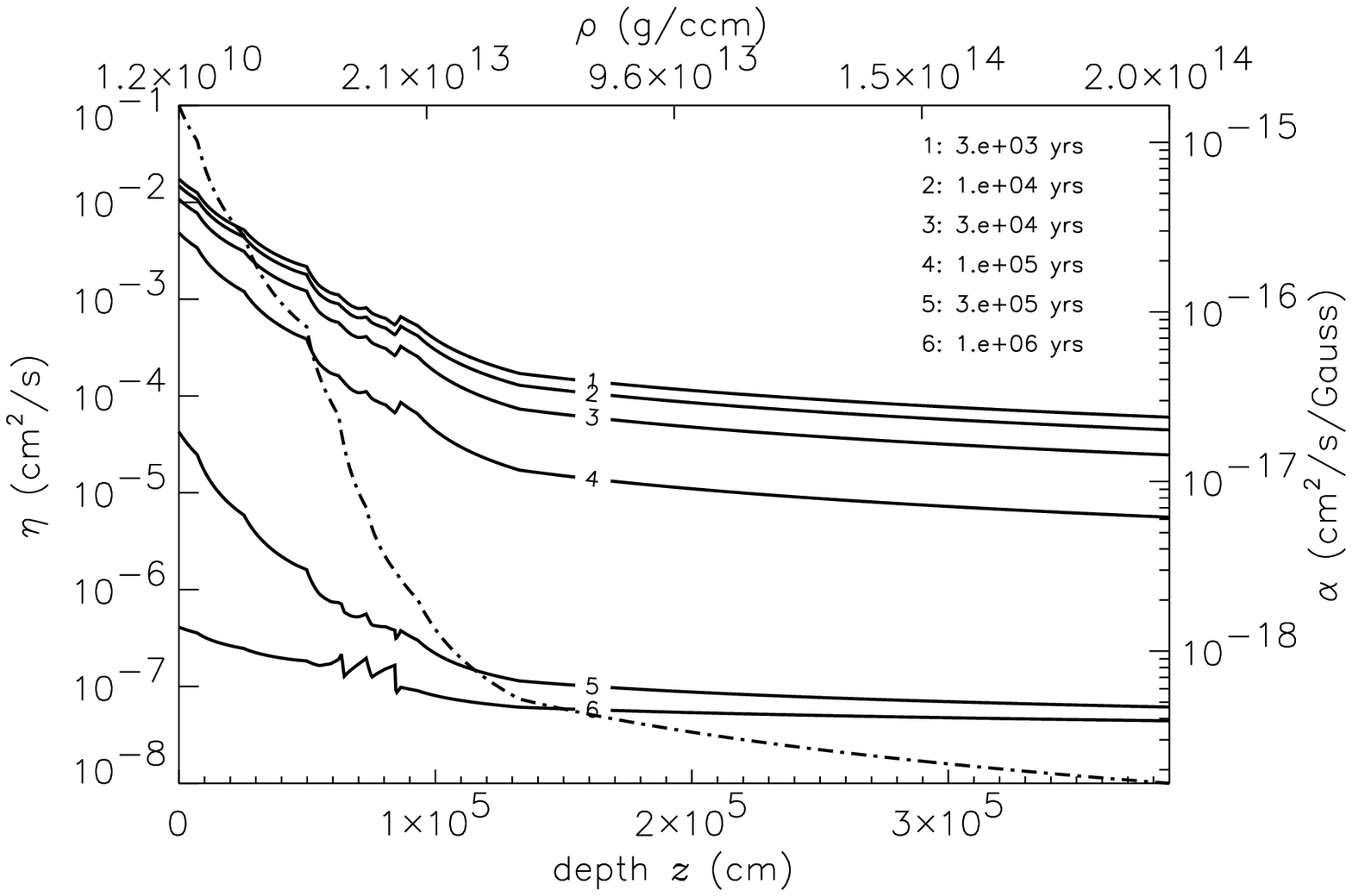, width=\linewidth}
\caption{\label{diffprofs}Radial profiles of diffusivity $\eta$ (solid) for different ages of the NS, and Hall coefficient $\alpha$ (broken). Top: FP, bottom: PS model.
All $\eta$ profiles for ages $\ge 10^7$ yrs (FP) and $\ge 10^6$ yrs (PS) practically coincide.}
\end{figure}

Finally, \figref{diffprofs} gives the profiles of the magnetic diffusivity $\eta$ and the Hall parameter $\alpha$ 
for both models.
For $\eta$ different profiles are shown corresponding to the ages at which the stability analysis will be performed,
but only one profile for $\alpha$ since it doesn't depend on temperature and hence not on age.
For $\eta$ an impurity concentration of 1\% was assumed.
Note, that in contrast to the PS model merely the implicit influence
of $A$ and $Z$ on $\eta$ and $\alpha$ via
the density is included for the FP model, the explicit one being neglected
as well as the influence of $X_n$.
Since the cooling of a NS 
proceeds faster in the PS model, further cooling will not affect the conductivity after about $5 \times 10^5$ yrs in this case.
Then phonons are practically no longer excited in the crustal crystal and
the conductivity is determined by electron--impurity collisions alone.
Hence, the diffusivity doesn't change any more, while for the
FP--model
%having a more than one order of magnitude longer typical cooling time,
a significant number of phonons may be excited up to an age of
approximately $5 \times 10^6$ yrs; therefore $\eta$ depends on time 
until this age.

According to the discontinuous behavior of $A$ and $Z$ in the PS model (see \figref{rhoprofs}), a non--smooth behavior
of its conductivity parameters appears in \figref{diffprofs}.
We decided not to
smooth out these discontinuities because it had meant introducing unnecessarily some further arbitrariness
into the model.

\subsection{Background field}
As the second major ingredient of a more realistic crust model
we have to specify background field profiles more justified with
a view to NS physics than
the ad--hoc assumptions employed in our former papers.
That's why we turn at this point to a model simulating the Ohmic decay of the
magnetic field in a cooling NS's crust (i.e., in a spherical shell), which is just the situation where we expect the
 most
 significant effects of the Hall--instability to occur. When thinking
about initial conditions for such simulations, one has to cope with the fact that,
unfortunately, there is less certainty about the very origin of the NSs' magnetic fields. Favorite mechanisms of magnetic field generation are 
proto--NS dynamos \citep{TD93} and thermoelectric instabilities \citep[see, e.g.,][]{WG96}, but
inheriting the magnetic field from the NS's progenitor seems well possible, too. Since the way of
field generation is surely significant with respect to the details of the field
structure, there seems to be no other resort  
when having to specify it than the principle of greatest possible simplicity. Therefore, we decided to consider dipolar fields only.
Further on,
it seems to be reasonable to assume with respect to their radial profiles that not all parts of the (in the case of a
proto--NS dynamo: later) crust equally participate in the field generation. Another process to be taken
into account, is the fallback accretion burying the magnetic field again within the crust soon
after its emergence.

On the other hand and somewhat conflicting with maximum simplicity, we had to ensure that enough curvature
is `injected' into the radial magnetic field profile initially. See \citet{RG02} for the essential role of the second derivative of the
background field's profile (which is partly taken over by the gradient of $\alpha$ in the present case as discussed above). 

At this point it is necessary to discuss the relationship between the
only $z$--dependent $\Bvec_0$ profile used in the plane slab geometry,
and a both $r$-- and $\theta$--depending profile resulting from decay calculations in a spherical shell.
Since $\Bvec_0$ as assumed in \eqref{b0} must be parallel to the boundaries, 
the plane slab model is only meaningful in the close vicinity of
the magnetic equator $\theta=\pi/2$ ($\theta$ -- polar angle).
$f(z)$ for a certain instant $\tau$ has therefore to be identified with the $\theta$--component of $\Bvec(r,\theta,\tau)$ where $r=R-z$,
$\theta=\pi/2$.
Accordingly, the initial background field is completely specified in geometry and strength
after having prescribed $f(z)$ for $\tau=0$, which we denote as $f_\tin$.
(The restriction to background fields parallel to the slab boundaries can be relaxed to 
fields with a constant vertical component. We will treat this case in a forthcoming paper.)

In view of all these aspects,
we generally chose initial profiles showing at least one maximum inside the crust and being zero beneath
a certain initial penetration depth $z_\tin$. %For the models considered
We fixed the latter by prescribing the corresponding density $\rhoinit$ to be either
$10^{13}$ or $10^{12} \gcmc$. The smaller $\rhoinit$ results in larger derivatives of the background field profiles
thus possibly favoring the instability at early instances. On the other hand the decay will be accelerated and
we have to expect smaller growth rates at higher ages (see Sect. \ref{results}).          
The following initial
profiles of the 3rd and higher degrees were chosen:
\begin{alignat}{3}
\hspace*{-2mm}f_\tin(z) &= -2 &&\Bd\!\left(\frac{R}{z_\tin}\right)^{\!2}\!\!\frac{z}{R-z}       \hspace{6.5mm}\,(1-\left(\frac{z}{z_\tin}\right)^{\!2}\!) &&\,\hspace{1mm}\text{cubic}\hspace{-1mm}\label{quart}\\
\hspace*{-2mm}f_\tin(z) &= -4 &&\Bd\!\left(\frac{R}{z_\tin}\right)^{\!4}\!\!\frac{z^3}{R^2(R-z)}\,(1-\left(\frac{z}{z_\tin}\right)^{\!4}\!)\; &&\,\text{heptic}\hspace{-1mm}\label{oct}\\
\hspace*{-2mm}f_\tin(z) &= \phantom{-}\pi &&\Bd  \,\frac{R}{z_\tin}\,\frac{R}{R-z}\,\sin\left(4\pi\frac{z}{z_\tin}\right) &&\hspace{-5mm}\text{sinusoidal}\hspace{-1mm}\label{cos}
\end{alignat}
Here, $\Bd$ 
is just the initial polar surface magnetic field.
Note, that the denominator $R\!-\!z$ in \eqref{quart}--\eqref{cos} doesn't influence
the degree of $f_\tin$ with respect to $z$ significantly since inside the crust $R\!-\!z\approx R$.

The decay calculations are straightforward \citep[for details see][]{PGZ00};
we only mention that the back--reaction of the magnetic field
on the temperature distribution and cooling history via Joule heating was neglected. Hence, the results depend
linearly on the initial conditions. 
The same boundary conditions
as referred to for the magnetic field perturbations were used. However,  
we refrained from adjusting the initial profiles \eqref{quart}--\eqref{cos} to the
vacuum boundary condition at the surface since the decay simulation 
will do this job automatically. (Likewise, diffusion smoothes out the `kink' at $z=z_\tin$.)
All runs were performed with $\Bdeins$ and
snapshots of the decaying field were taken at nine instants from $3\times 10^3$ to $3\times
10^7$ yrs in order to define the profiles $f(z)$.
Additionally, we scaled them (that is, $\Bd$) by factors 2 or 5 to study the
dependence of the instability on the initial 
field strength.

\begin{figure}[H]
\begin{center}
\epsfig{file=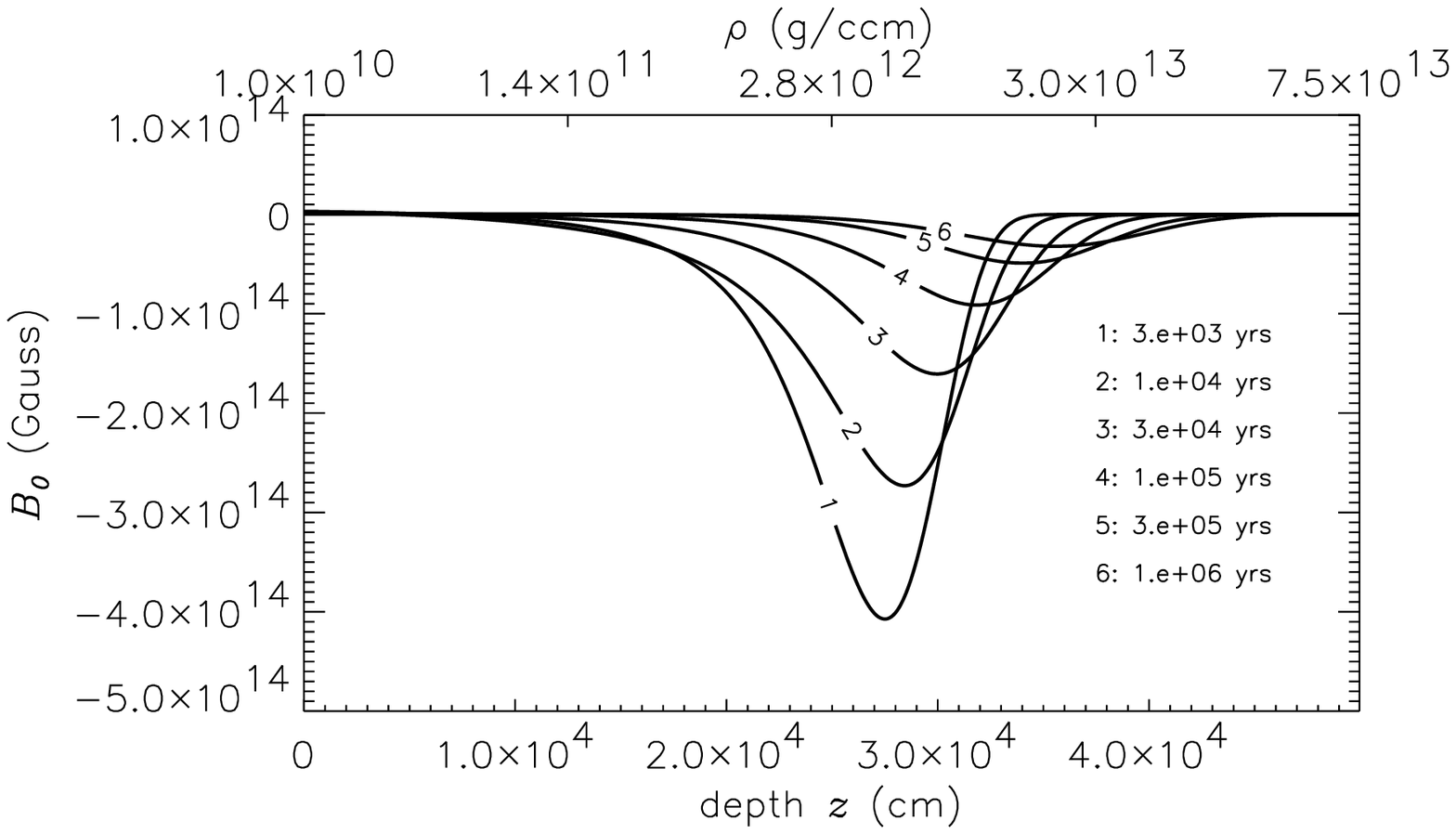  , width=.9\linewidth}  \\
\epsfig{file=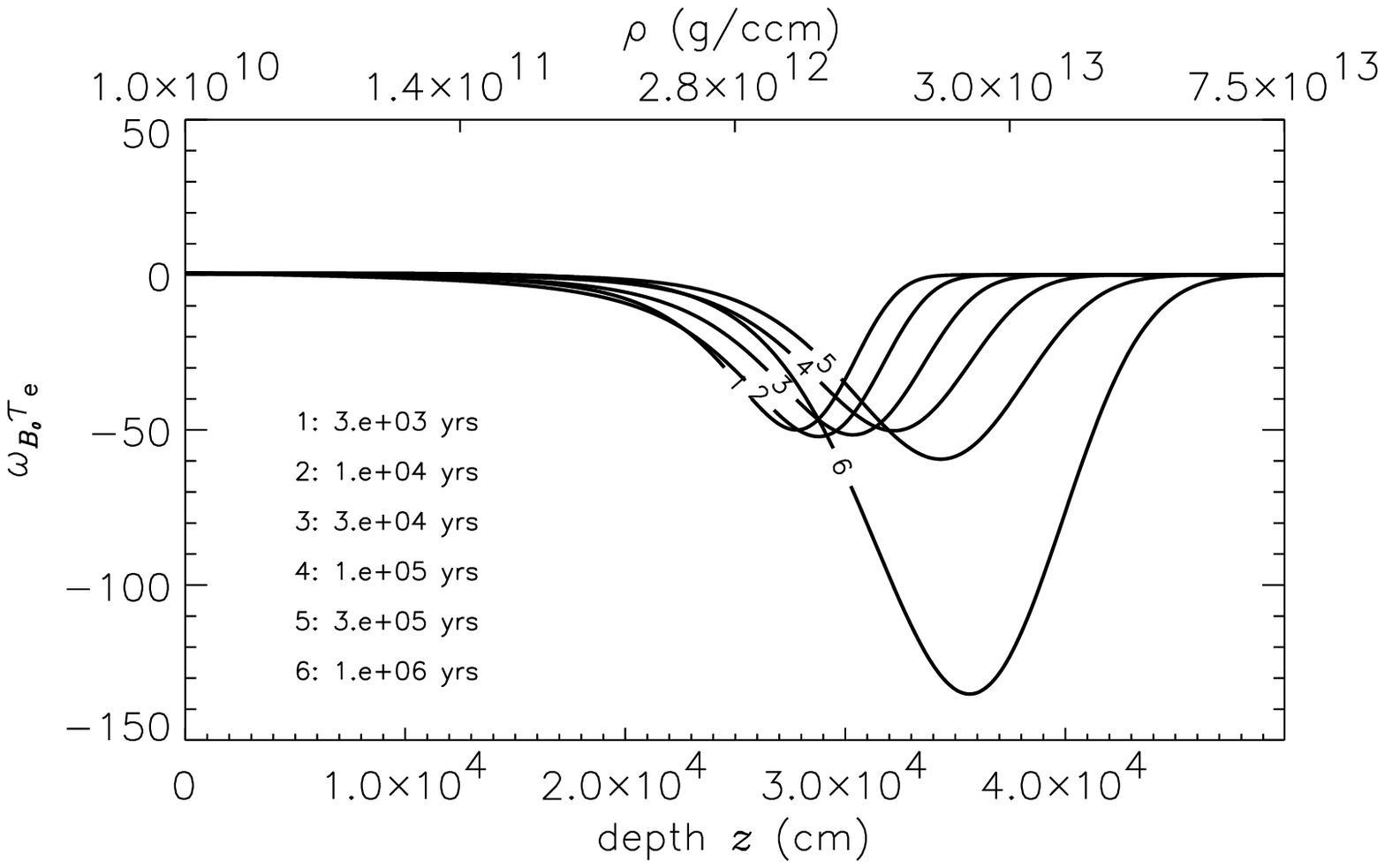, width=.9\linewidth}  \\
\epsfig{file=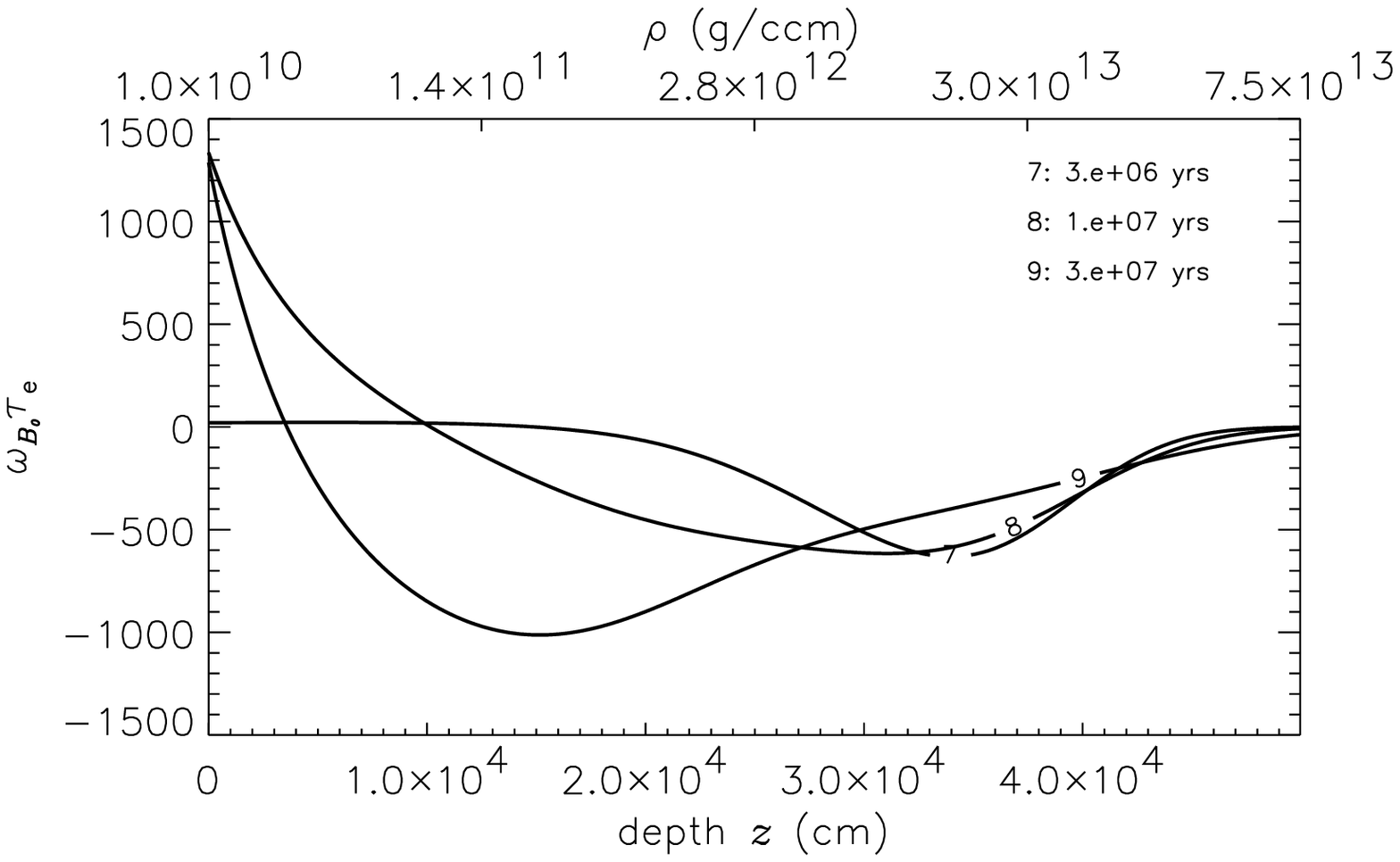, width=.9\linewidth}  \\
\hspace*{-.04\linewidth}\epsfig{file=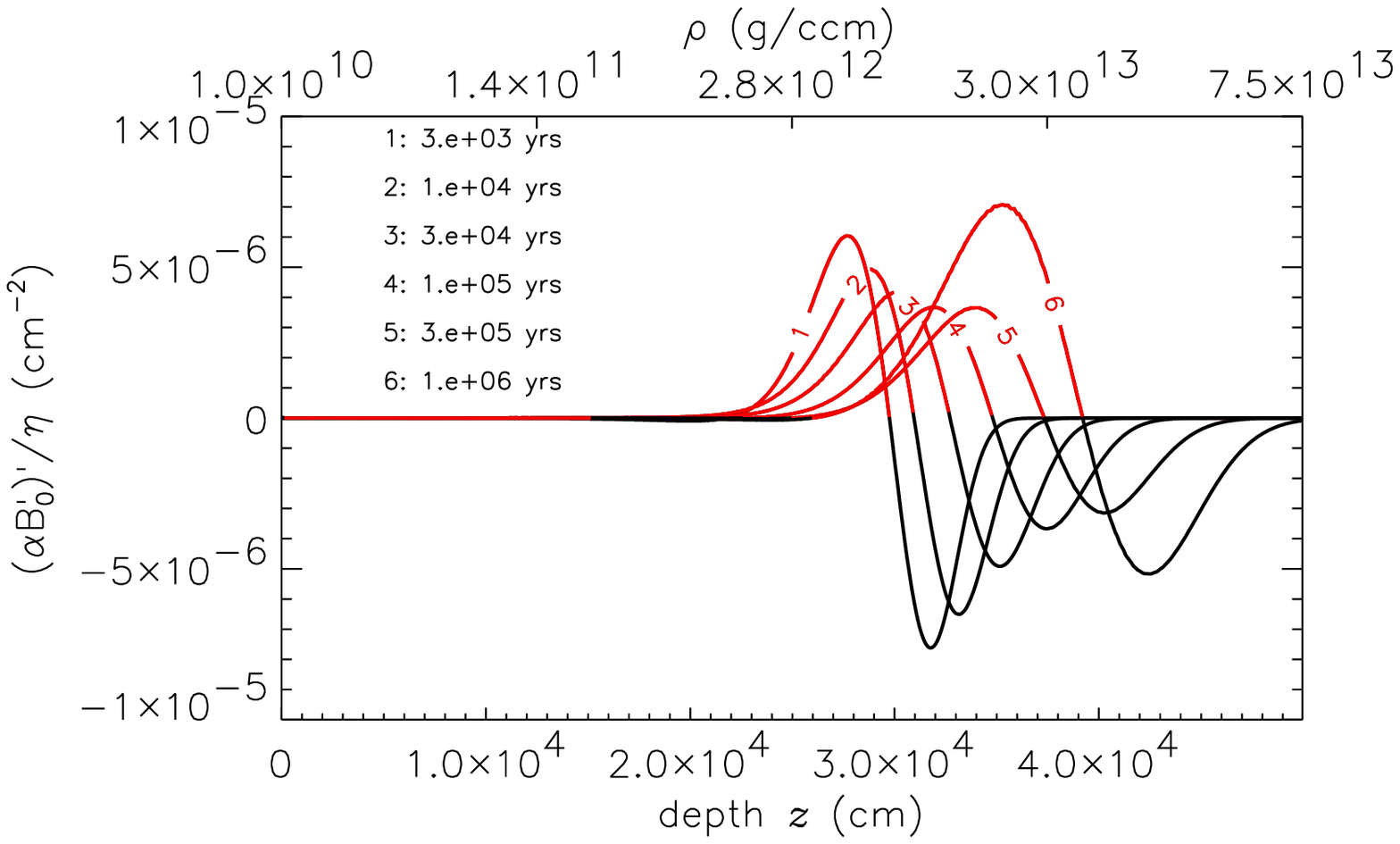, width=.94\linewidth}\\
\hspace*{-.04\linewidth}\epsfig{file=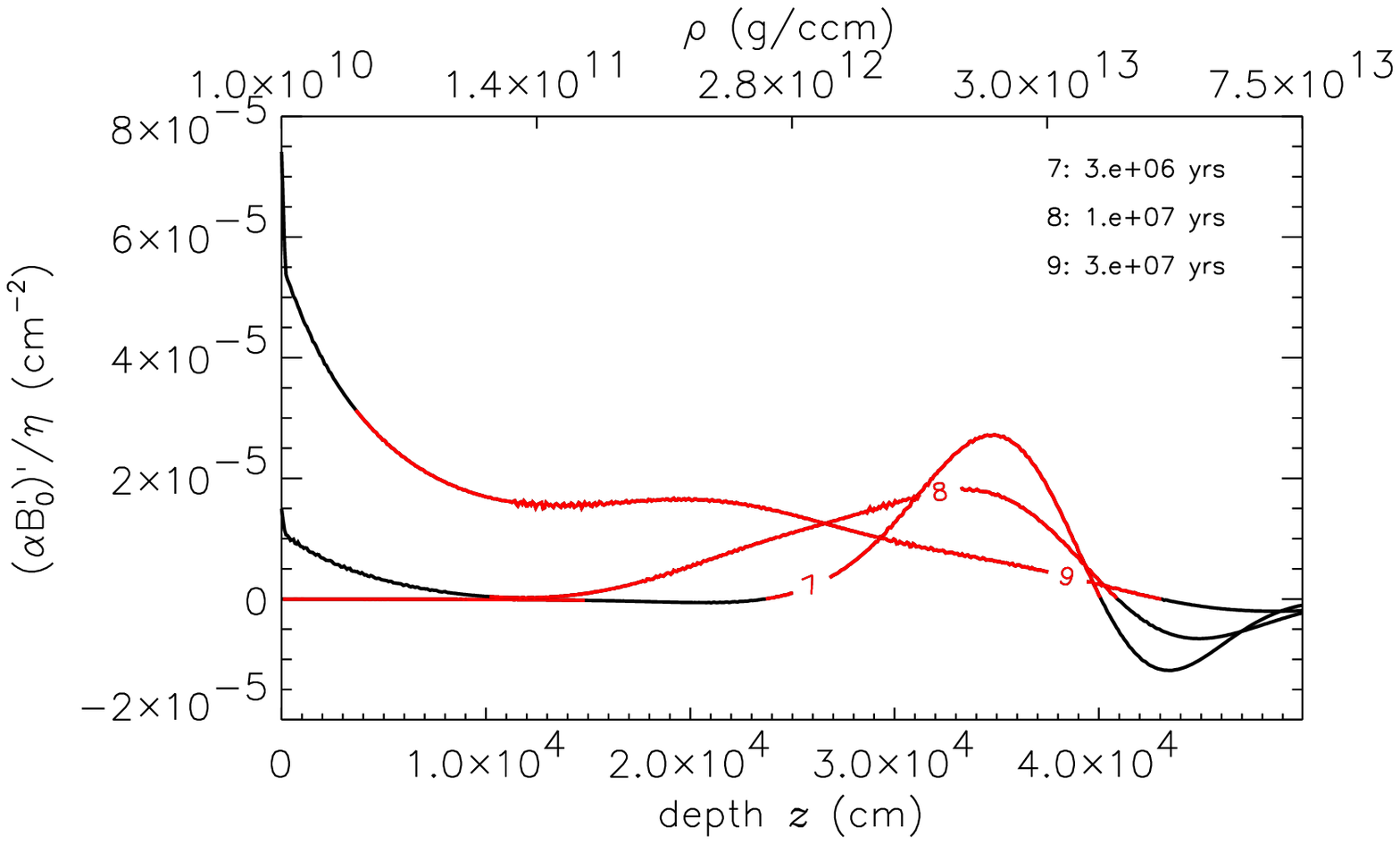, width=.94\linewidth}
\end{center}

\vspace{-3mm}
\caption{\small\label{FPcprofs}Background field $\Bzero$, magnetization parameter $\ombzert$ and 
curvature parameter $(\alpha\Bzero')'/\eta$ resulting
from the FP model with sinusoidal initial profile \eqref{quart} and
initial penetration density $\rhoinit=10^{13} \gcmc$, and $\Bdeins$. Red parts indicate adherence to the sign condition
\eqref{signcon}.}
\end{figure}

\begin{figure}[H]
\begin{center}
\epsfig{file=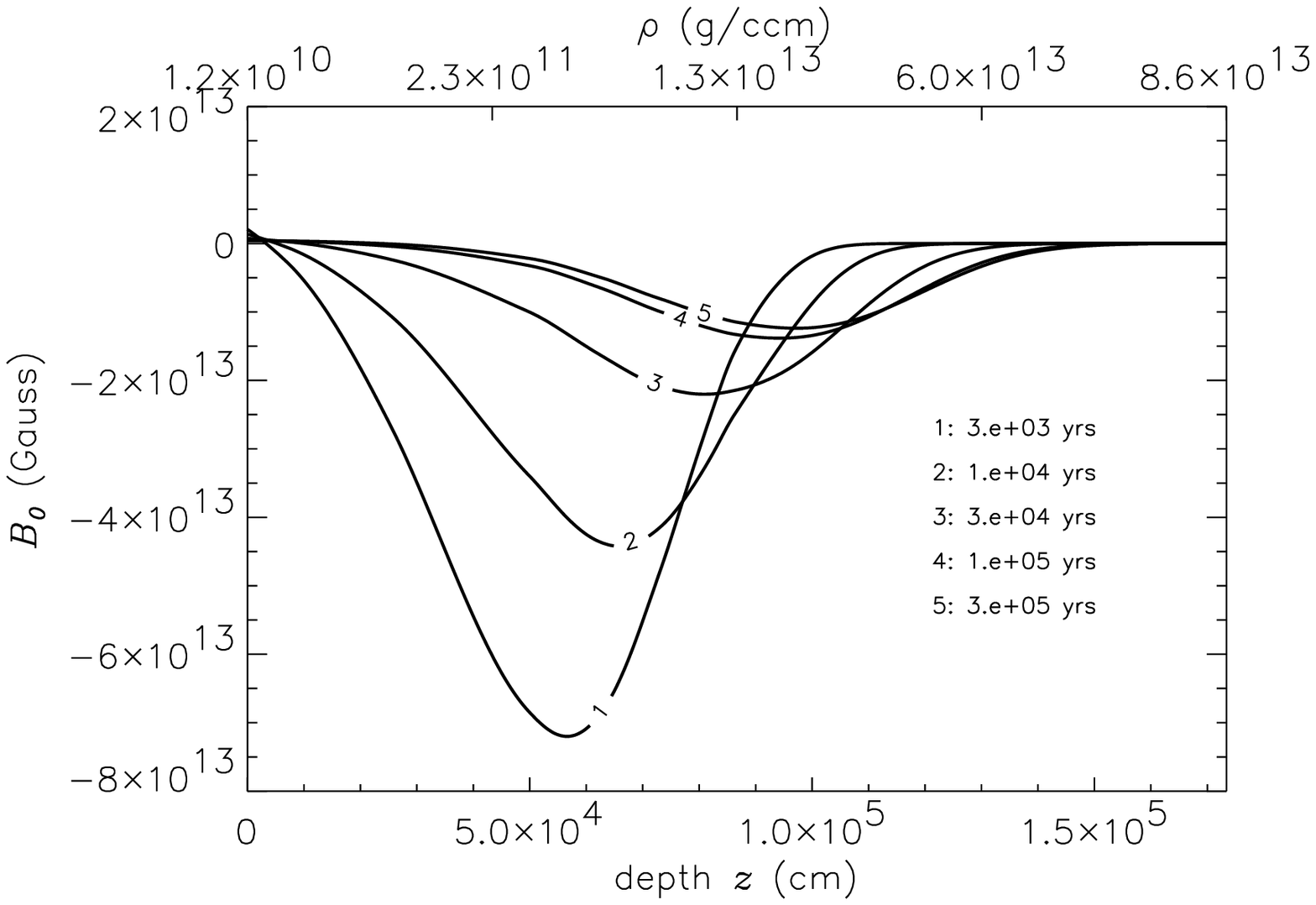, width=.93\linewidth}
\epsfig{file=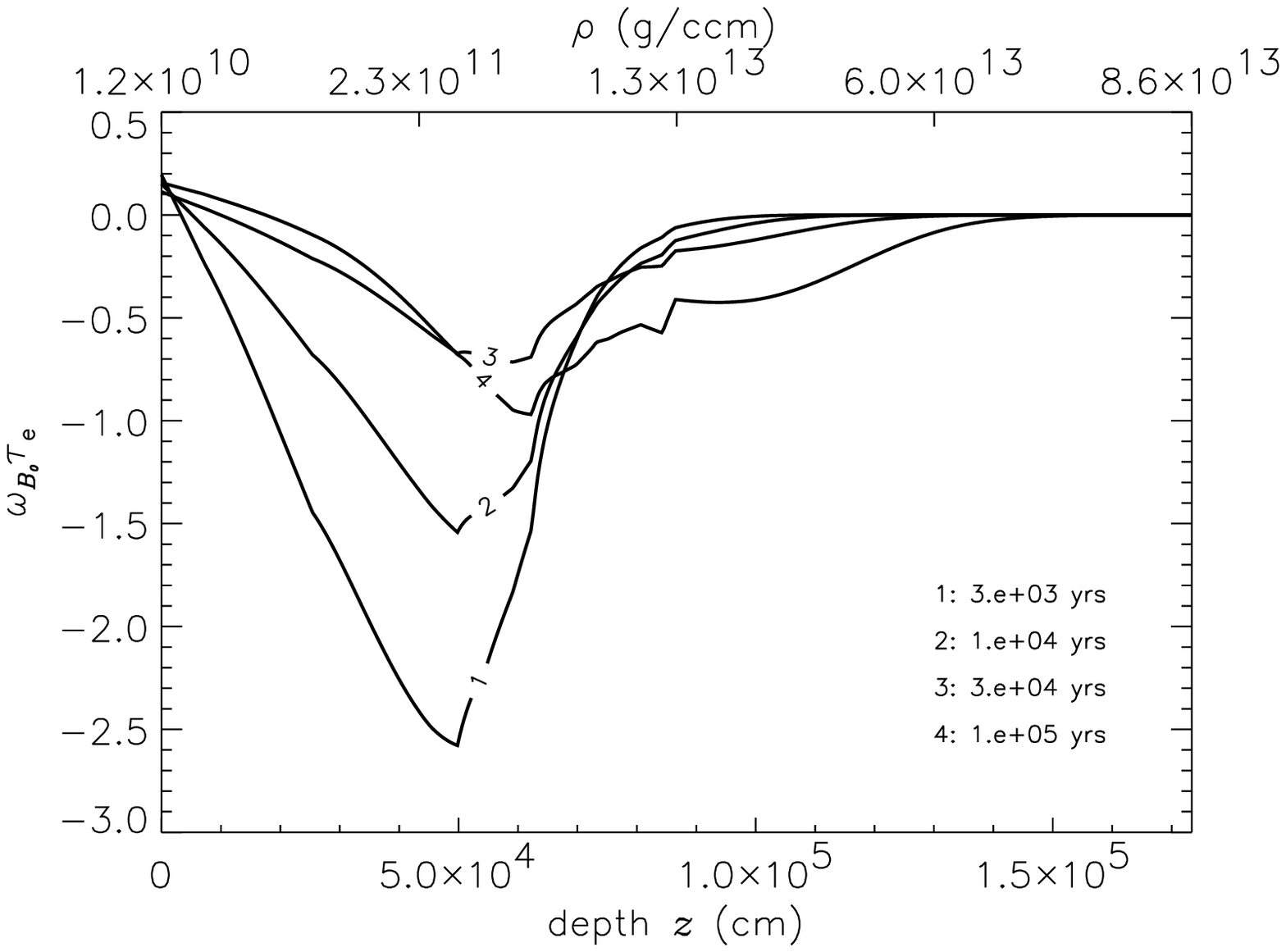, width=.93\linewidth}
\epsfig{file=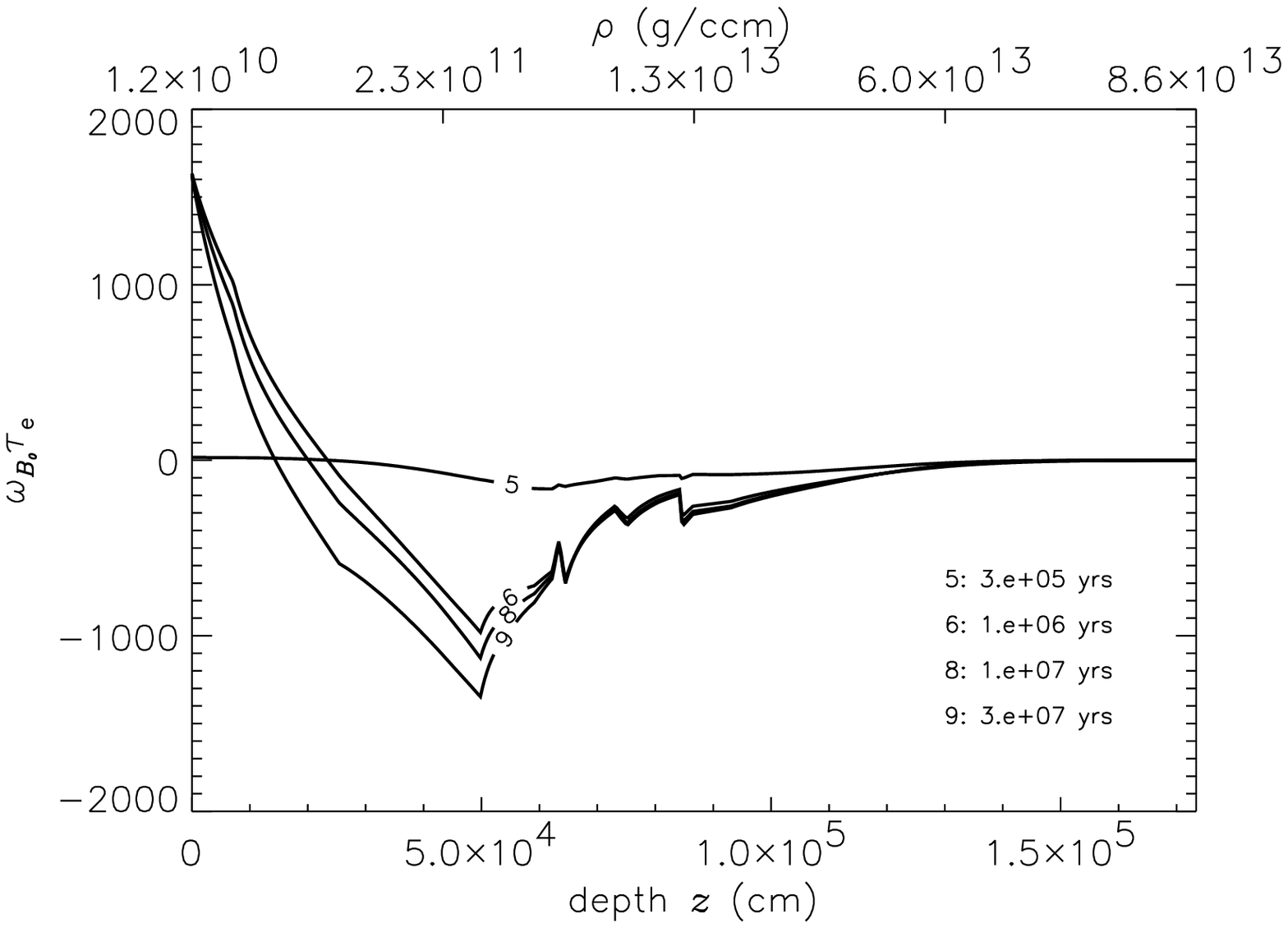, width=.93\linewidth}
\end{center}

\vspace{-.2cm}
\caption{\label{PSqprofs}Background field $\Bzero$ and magnetization parameter $\ombzert$ resulting
from the PS model with cubic initial profile \eqref{quart} and 
initial penetration density $\rhoinit=10^{13} \gcmc$, and $\Bdeins$.}
\end{figure}

\begin{figure}[t]
\epsfig{file=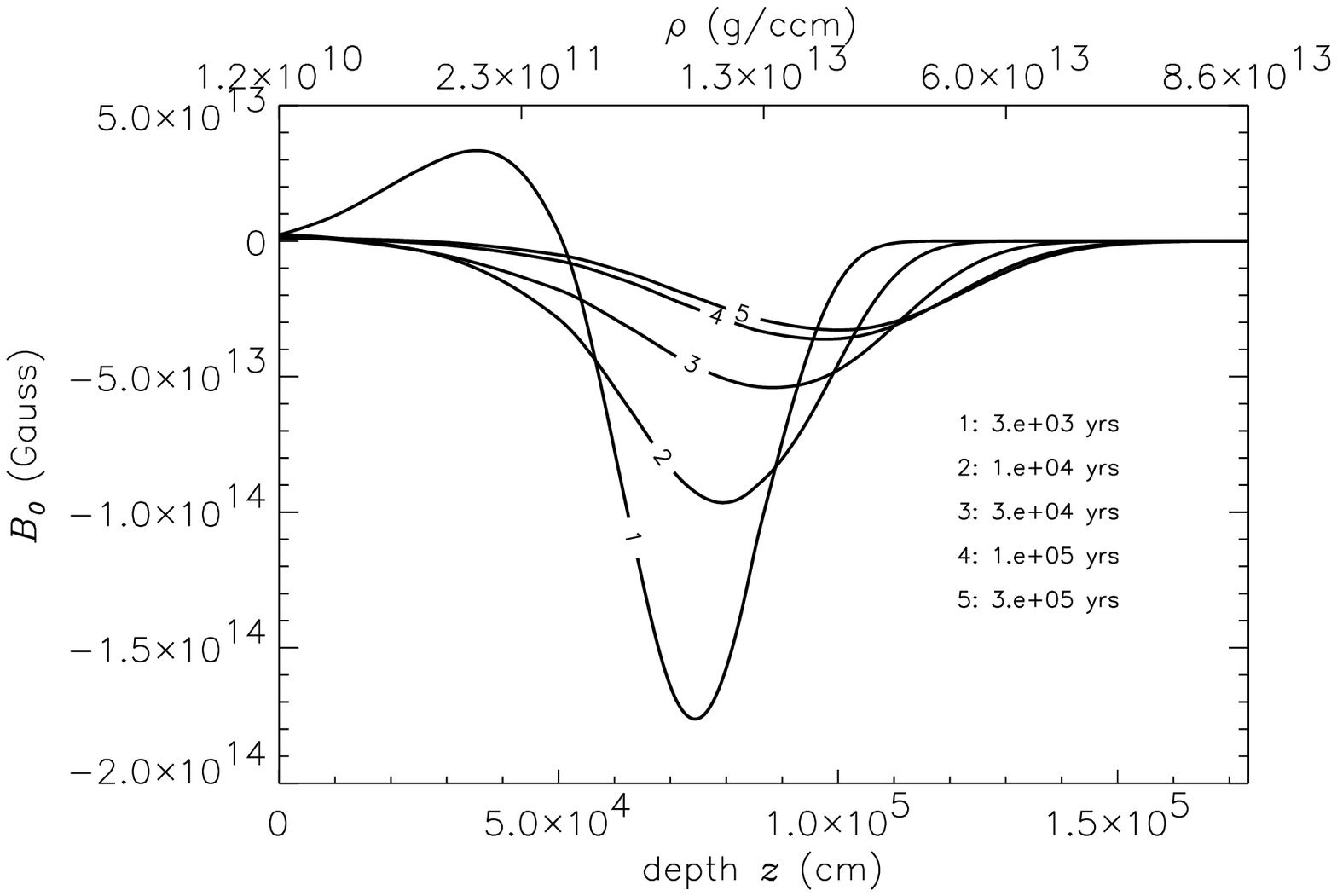, width=.93\linewidth}
\epsfig{file=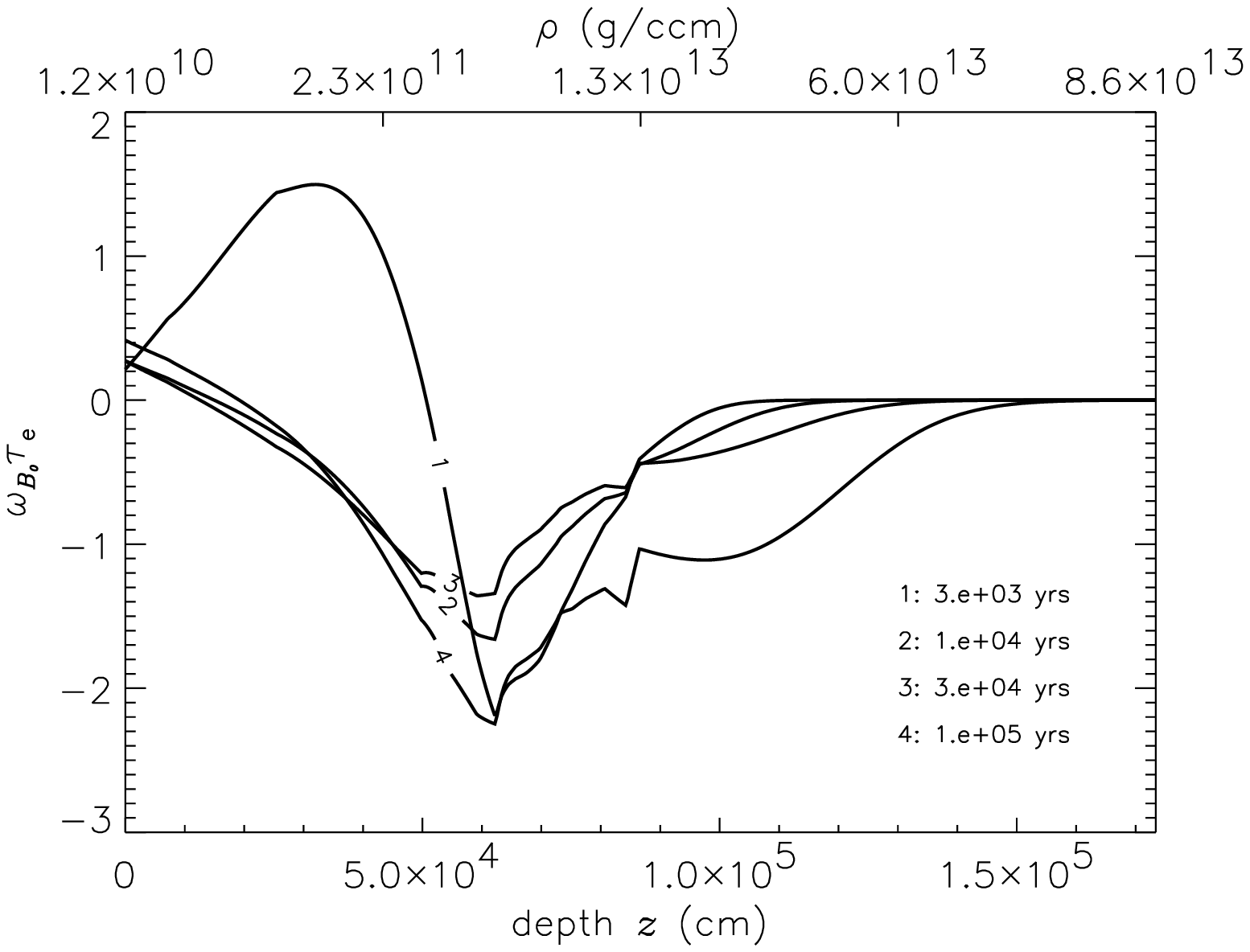, width=.93\linewidth}
\epsfig{file=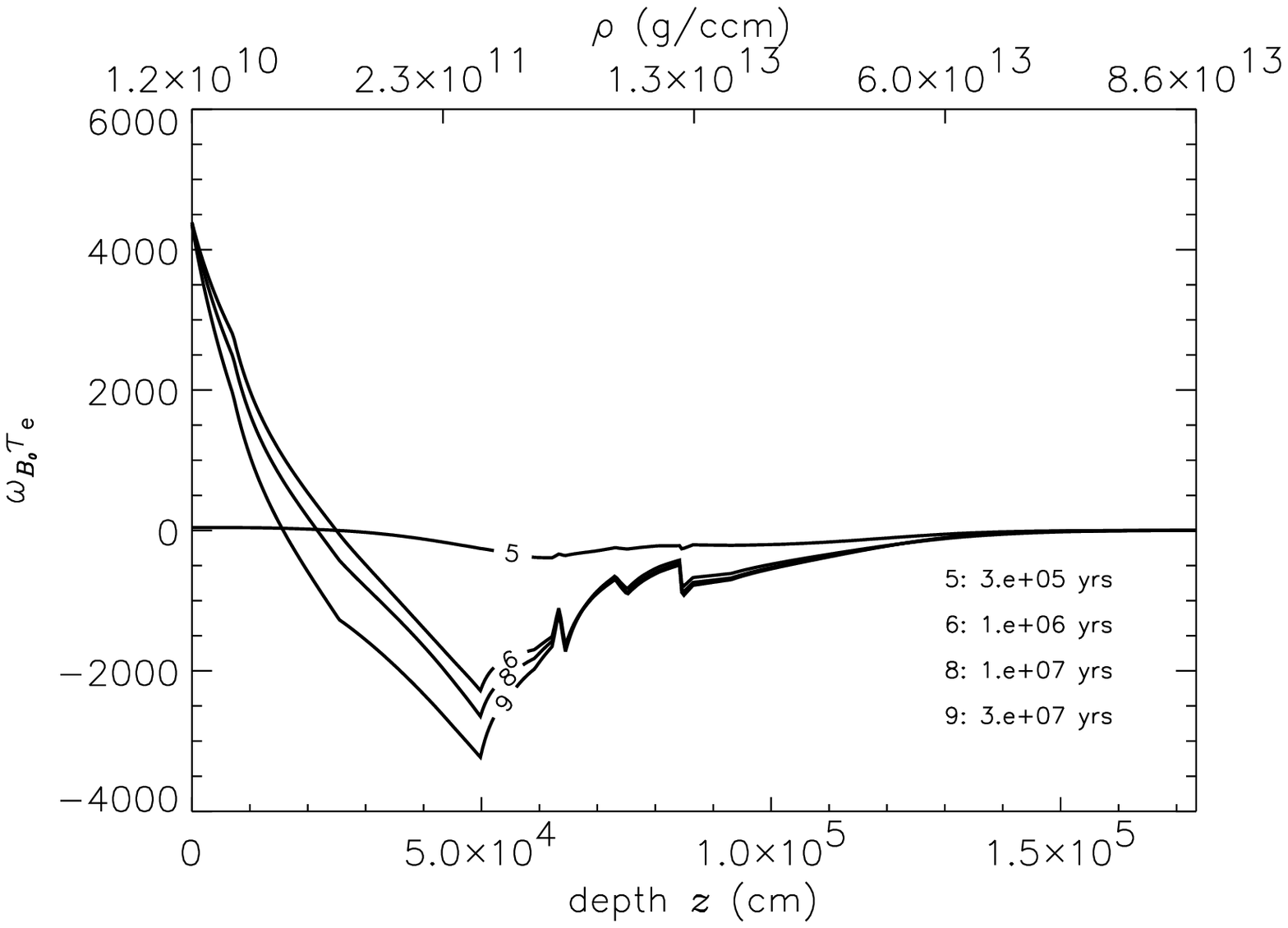, width=.93\linewidth}
\caption{\label{PScprofs}As \figref{PSqprofs}, but with sinusoidal initial profile
\eqref{cos}.}
 
\vspace{-1mm}
\end{figure}

Figures~\ref{FPcprofs} to~\ref{PScprofs} show typical examples of the resulting $f(z)$ for both models.
(See also \figref{geometrie} for schematic sketches of characteristic
background fields.) Moreover, these figures present the magnetization parameter $\ombzert$ and 
--- only for the FP model, in \figref{FPcprofs} --- the
quantity $(\alpha f')'/\eta$, further on referred to as ``curvature parameter". 
\noindent
It is an estimate for the ratio of the potentially energy--delivering term $\curl(\alpha\curl\Bzero\times\bvec)$
to the (anyway) energy--dissipating
term $\curl(\eta\curl\bvec)$ in \eqref{indeqlin}. (Note that the
term $\curl(\alpha\curl\bvec\times\Bzero)$ is energetically neutral.) Therefore we suppose that the
curvature parameter rather than $\ombzert$ itself is the key indicator with respect to the
vigour of the
instability. For the PS model, the curvature parameter is
only piecewise continuous due to the non--smooth profiles of $\eta$ which result in $f''$ profiles only
piecewise continuous, too. Nevertheless, the solutions of \eqref{indeqst} are well defined, at
least in the mathematically weak sense.

On the other hand, we want to stress that the {\em magnitude} of the curvature parameter alone is again not
sufficient to infer the existence of the instability, which relies on a complicated interplay of
{\em vector} quantities. Therefore, an additional qualitative criterion must be satisfied in order
to have the potentially energy--delivering term in \eqref{indeqlin} {\em really} energy-delivering. From simple qualitative
considerations with axisymmetric fields in a homogeneous medium it is possible to conclude that
$f''(z) f(z) <0$
has to be satisfied, at least over a certain $z$-interval. Or generalized:
Background field and curvature parameter must have {\em different} signs:   
\begin{equation}
 (\alpha f')'(z) f(z) <0\;.
    \label{signcon}
\end{equation}

\section{Results}
\label{results}
\subsection{Numerics}
The system of ordinary differential equations \eqref{indeqst} was discretized by help of symmetric difference formulae
of second order; near boundaries unsymmetric formulae were employed, if necessary.
In most cases an equidistant grid with a typical node number of 1200 was used, but in quite a number of spot
checks of convergence the node number was repeatedly doubled up to 9600 maximum.

For the numerical solution of the resulting complex non-Hermitian standard matrix eigenvalue problem
we took benefit of the package ARPACK (routines {\tt znaupd, zneupd}).
Some results were checked applying the LAPACK routine {\tt zgeev}.
The results from both packages agree in at least six digits
(but comparisons were possible for moderate node numbers only).

Unfortunately, the order of the difference formulae was not always reflected by the convergence rate
of the eigenvalues with respect to the node number:
For the PS model convergence was usually significantly slower.
Perhaps this behavior is connected with the  
discontinuities of the coefficients $\eta'$, $\alpha'$ and $f''$ described above.
This supposition is supported by the fact that the convergence rates of the FP
model are close to (although not exactly) quadratic.

\subsection{General aspects of the results}
As with the homogeneous density model, both oscillating and non-oscillating unstable modes
exist, where the fastest growing mode of any specific model turned out to be always amongst the latter.
As another common feature of the fastest growing modes, their wavenumber $k_y$
has always been found to be zero.
When deriving the background field from an axisymmetric poloidal one in the vicinity of its magnetic equator
(as we did), association of the $y$--co--ordinate with longitude is surely the proper choice.
Thus, with some care one may suppose that in a spherical shell the most unstable modes are preferentially axisymmetric.
\begin{figure*}[t]
\begin{tabular}{cc}
\epsfig{file=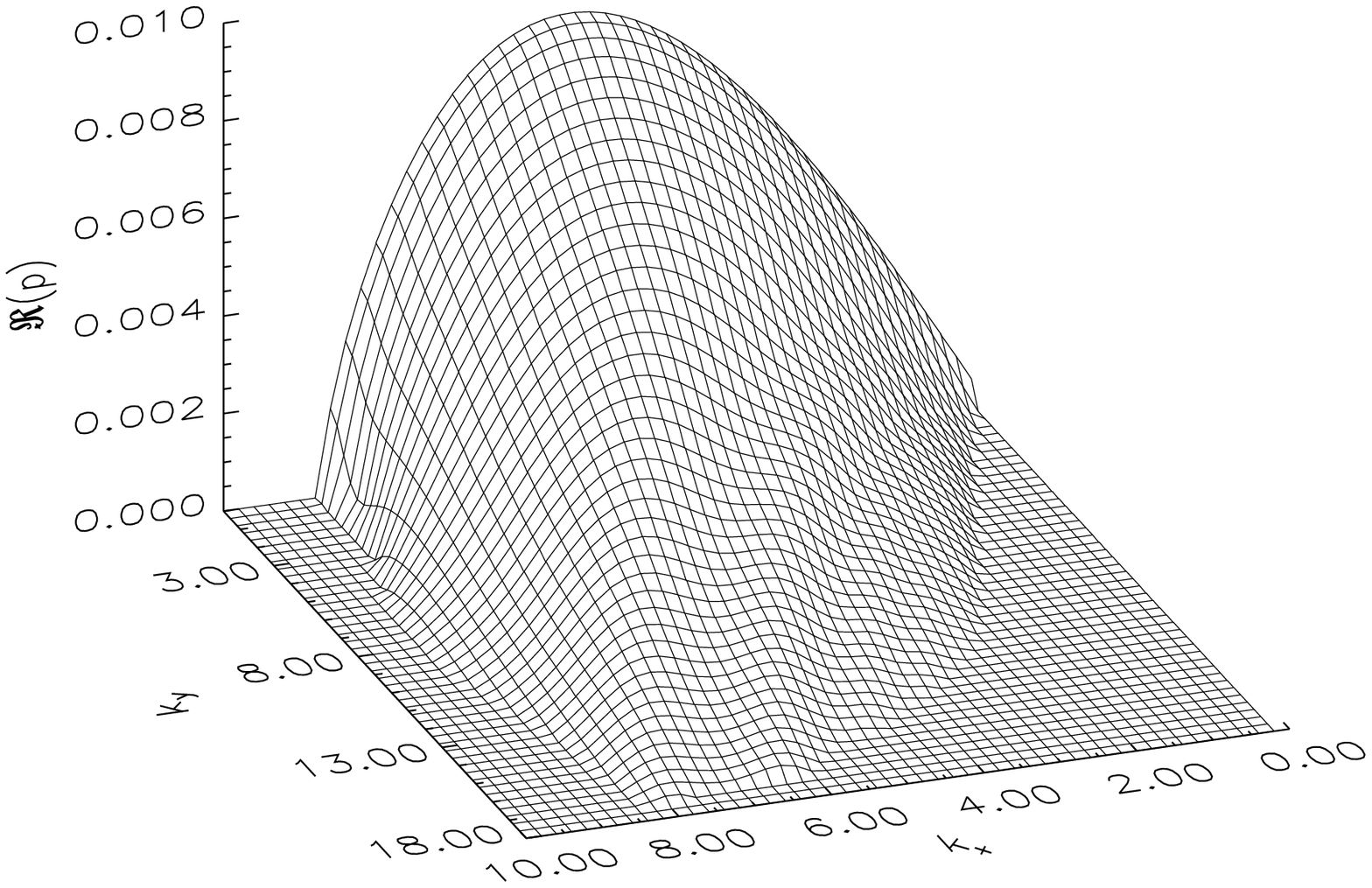, width=.45\linewidth}&
\epsfig{file=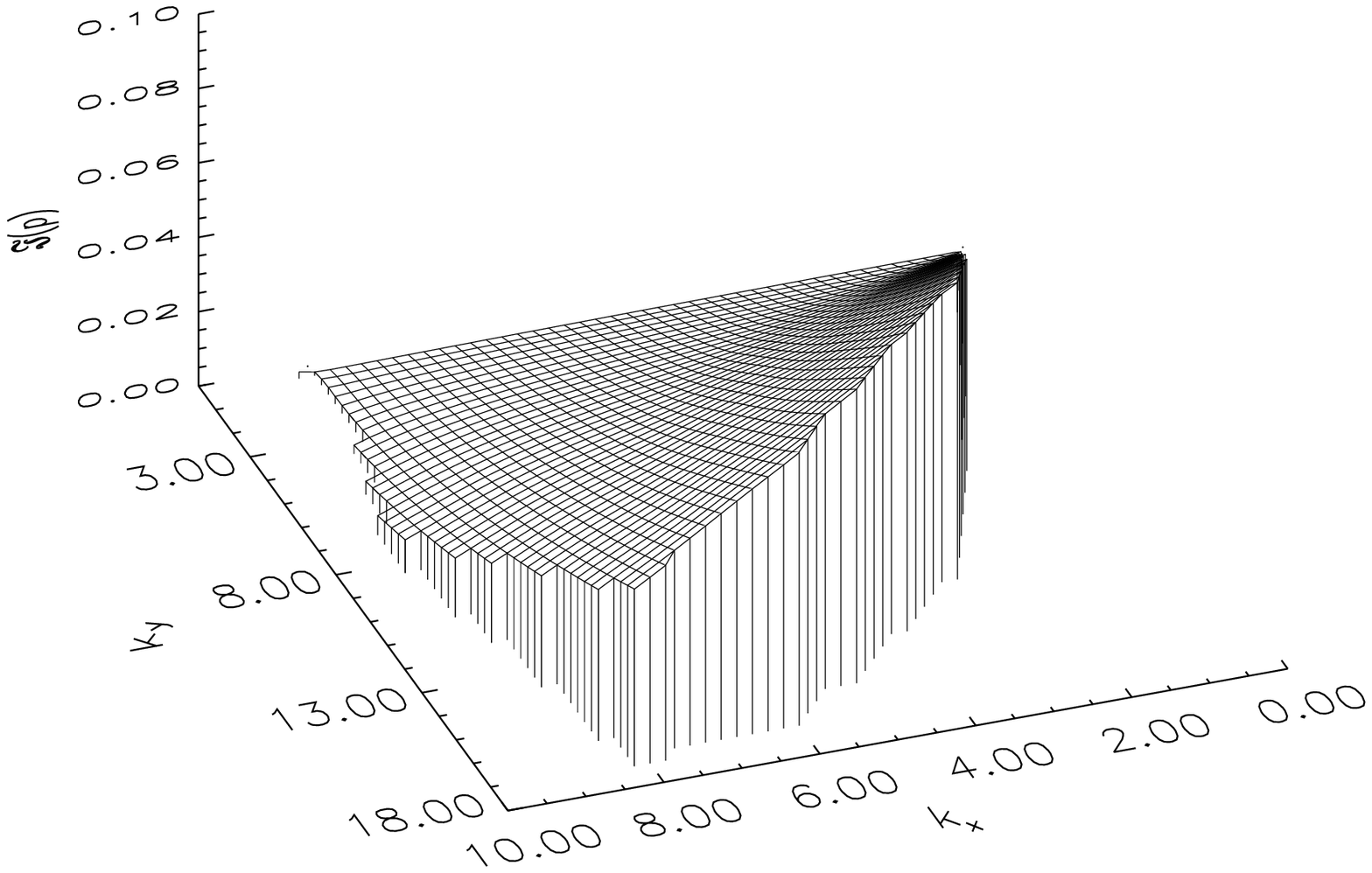, width=.45\linewidth}
\end{tabular}

\vspace{-3mm}
\caption{\label{gebirge}Eigenvalue $p$ as function of the wave numbers $k_x$, $k_y$ for the PS model with cubic
initial background field profile \eqref{quart}, initial field $\Bdfuenf$,
initial penetration density $\rhoinit=10^{13} \gcmc$, age $3\times 10^5$ yrs. Left: growth rate $\Re(p)$, negative values were set to zero;
right: frequency $\Im(p)$, not shown for $\Re(p)<0$. $p$ in units of $4/(d/\text{cm})^2 \text{s}^{-1} = 2.68\times 10^{-11} \text{s}^{-1}$, wave numbers 
in units of $2/d= 5.18\times 10^{-6} \text{cm}^{-1}$.}
\end{figure*}

Figure~\ref{gebirge} shows growth rate $\Re(p)$ and oscillation frequency $\Im(p)$
as functions of $k_x$ and 
$k_y$ for a typical example (PS model, age $=3\times 10^5$ yrs, cubic initial field, $\Bdfuenf$
, $\rhoinit=10^{13} \gcmc$).
The rectangle enclosing the 
unstable region in the $k_x$--$k_y$--plane is defined
by $0.4/d\le k_x\le 18.4/d$, $0\le k_y\le 34.4/d$.
But, when obeying the background dynamics permissibility condition (see Sect. \ref{baseq})
only a considerably smaller unstable region can be considered feasible.

In contrast to the growth rates, the oscillation frequencies grow in general with $k_y$; their
highest values seem to occur always at the edge of the unstable region in the
$k_x$--$k_y$--plane, that is, for $\Re(p)\rightarrow 0$ (and $k_y\ne 0$).
Note, that the oscillating unstable modes can be regarded as helicoidal waves
with growing amplitudes. (Cf. \citet{VCO00} who considered damped helicoidal waves in a stratified crust.)

In \figstoref{tiperqfp}{tipercps} the growth times of the fastest growing modes $1/\tmax(\Re(p))$
(simply referred to as ``growth time" $\taugrow$ of a specific model)
for the most important cases considered, are presented as functions of age and initial field strength $\Bd$.
In addition, the tangential period lengths of these
modes, $\lamax$, are given; because of $k_y^\tmax=0$ this quantity is defined as
\begin{equation}
\lamax = \frac{2\pi}{k_x^\tmax} \;. 
\end{equation}
It should be regarded as one of the dominating scales of the unstable modes since, in contrast
to the homogeneous density model, in the majority of cases no prominently small radial scales were found (see
\figref{eigfunc}). Instead, the radial scales of the unstable modes seem to be determined simply by
the radial scale of the background field.
Except for some earlier stages of the PS model, $\lamax$ depends in general only slightly on 
$\Bd$. Since the wavenumber $k_x$ was given only discrete equidistant values,
the period lengths for different values of $\Bd$ frequently even
coincide (indicated in the figures by filled symbols instead of open ones).
\subsection{Growth times}
We stress again that the growth times have to be considered as referring to
snapshots only. That is, the value for a given NS age has been calculated
assuming that the instability starts just at that age, with the background field
given at exactly that moment.
Clearly, the results must not be interpreted as a sort of
temporal evolution of the growth times: Any occurrence of the
Hall--instability will affect the strength and structure of the background field,
thereby changing the conditions for the occurrence of the instability itself at
later moments. This shortcoming of the results induces us to emphasize once more
the necessity of full non--linear calculations.

As an overall property, we state that the dependence of $\taugrow$ upon the initial
polar magnetic field $\Bd$ is always {\em monotonically falling}. Hence, we never entered the
range of $\Bd$ where higher values may yield larger growth times \citep[see][]{RG02}. 

\subsubsection{FP model}

\begin{figure}[t]
\epsfig{file=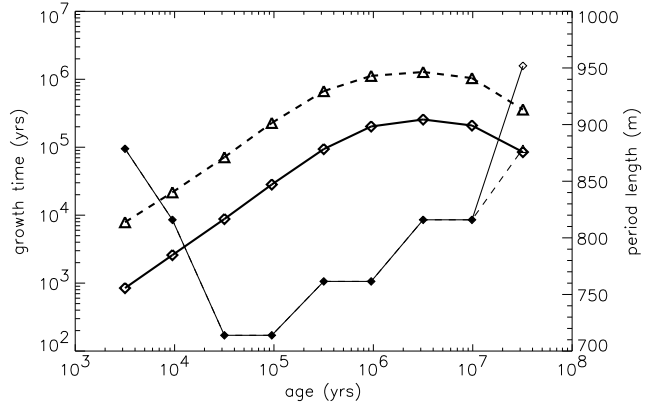, width=\linewidth}
\caption{\label{tiperqfp}Growth times and tangential period lengths of the fastest growing modes for the FP model with
cubic initial background field profile \eqref{quart},
initial penetration density $\rhoinit=10^{13} \gcmc$.
Solid, diamonds: $\Bdfuenf$; 
dashed, triangles: $\Bdeins$. Thick lines, big symbols: growth times; thin lines, small symbols: period lengths.
Full small diamonds: coincidence of the period lengths for different values of $\Bd$.}
\end{figure}

The growth times of the FP model (\figstoref{tiperqfp}{tipercfp}, \tabref{compFP1}) have in common that they start with
small values (most in the order of magnitude $10^3$ yrs and smaller)
at the youngest age, exhibit a nearly linear dependence on age (a power law with an exponent
between 0.8 and 1) until $10^5\ldots\,3\times 10^5$ yrs,
reach a maximum at $10^6\ldots\,3\times 10^6$ yrs, and fall progressively with age later on.
%We suppose, that the latter tendency even continues beyond the highest age considered ($3.2\times 10^7$ yrs).
At early stages the background dynamics permissibility condition
is well fulfilled for $\Bdfuenf$ but must be put in question in most cases with $\Bdeins$. 
Analogously, the maximum growth times are reliable for most cases with $\rhoinit=10^{13} \gcmc$, but
questionable for some of the cases with $\rhoinit=10^{12} \gcmc$ and $\Bdeins$. 
At the latest stage considered all the growth times satisfy the condition.
We note that the dependence of the growth time on $\Bd$ is not far from being linear for  
$\rhoinit=10^{13} \gcmc$, whereas for $\rhoinit=10^{12} \gcmc$ the dependence is of higher order. In this case,  
with the cubic profile and $\Bdeins$ even a ``gap" in the growth time curve occurs: From the age of $10^4$ to the age of $3\times 10^5$ years
there were no growing modes to be found at all. 

Comparing the growth times for different values of $\rhoinit$, one has to state that 
the smaller initial penetration depth ($\rhoinit=10^{12} \gcmc$), although being connected with
stronger gradients of the background field, is nevertheless disfavoring the instability. This tendency
becomes increasingly apparent with growing age. We explain it with the accelerated decay of these
``shallower" profiles.

Comparing the growth times for different initial profile types, one can see that the values for the heptic and 
the sinusoidal profiles are close together at the earlier stages
\enlargethispage{\baselineskip}
\begin{figure}[h]
\epsfig{file=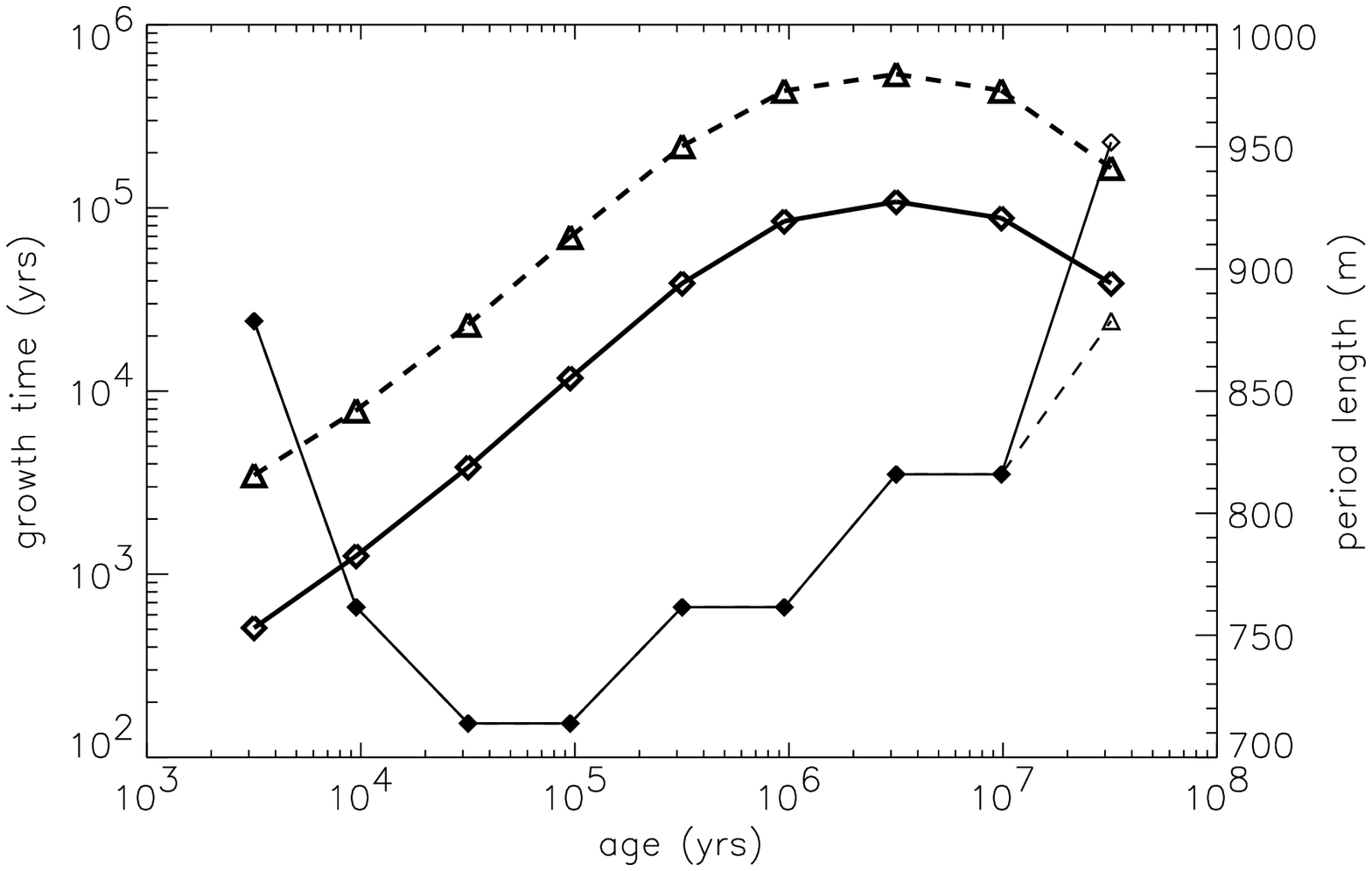, width=\linewidth}
\caption{\label{tiperofp}As \figref{tiperqfp}, but with heptic initial background field profile \eqref{oct}.}
\end{figure}
\begin{figure}[h]
\epsfig{file=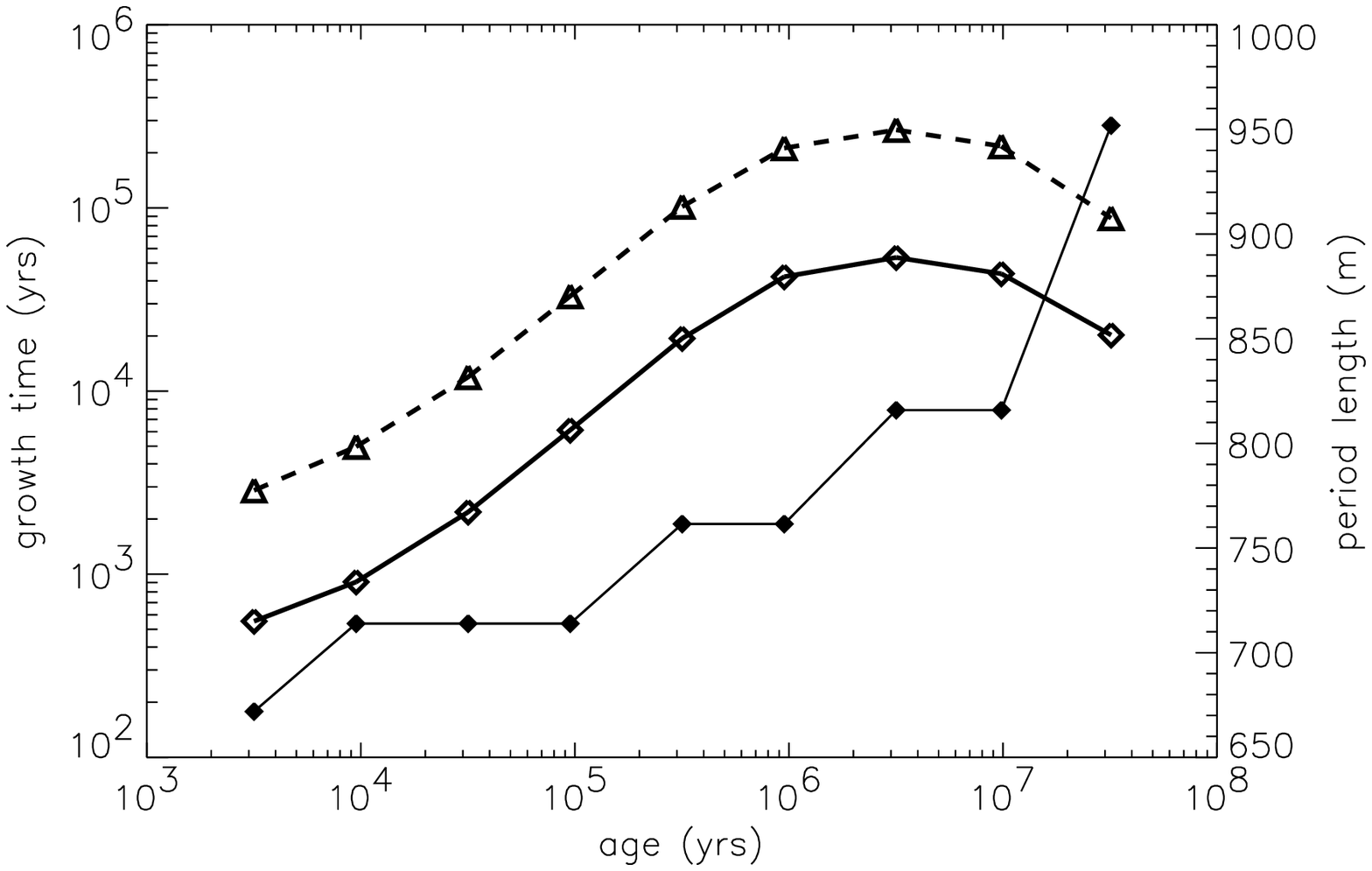, width=\linewidth}
\caption{\label{tipercfp}As \figref{tiperqfp}, but with sinusoidal initial background field profile \eqref{cos}.}
\end{figure}
\begin{figure}[H]
\begin{center}
\epsfig{file=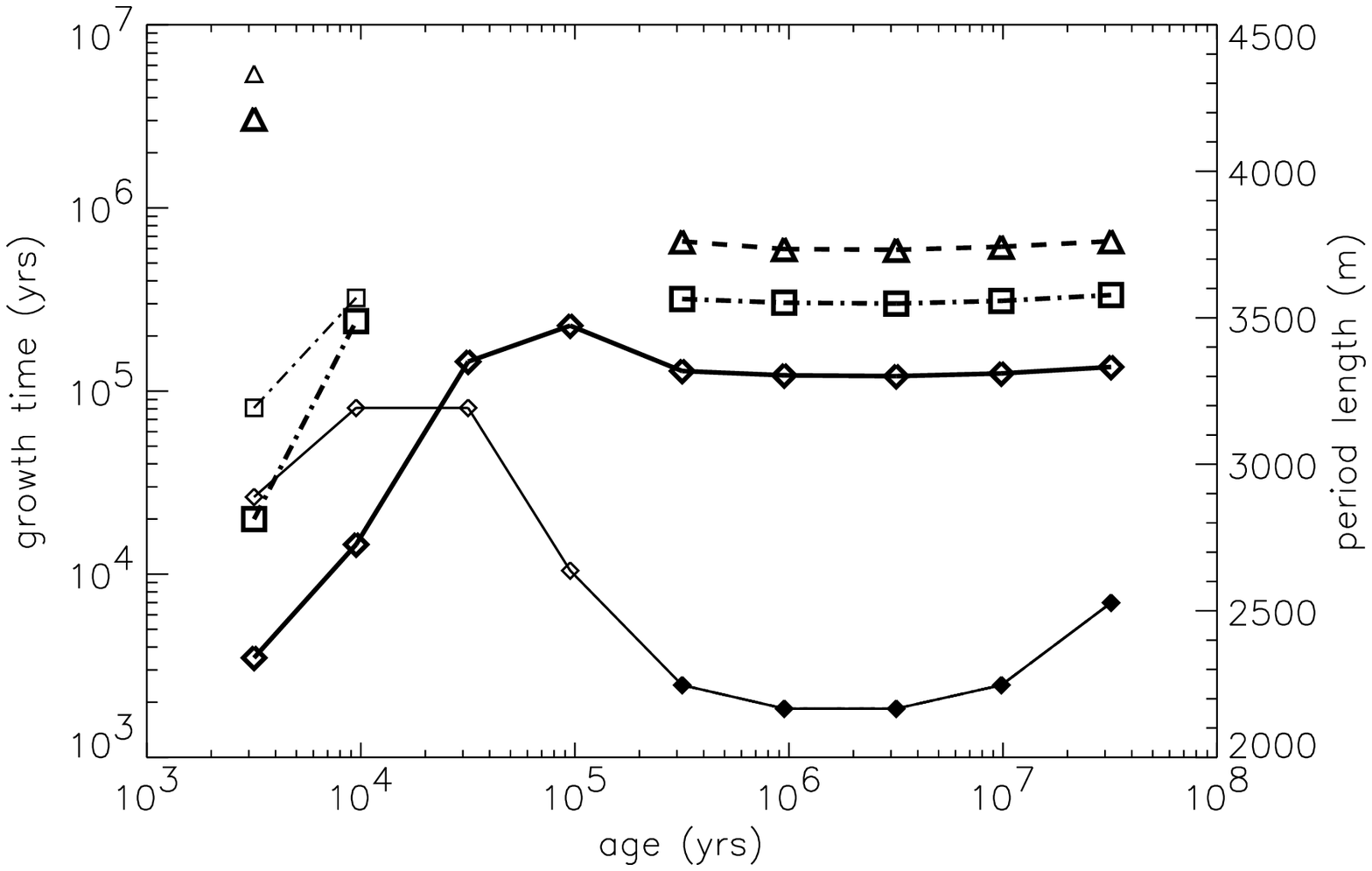, width=\linewidth}
\caption{\label{tiperqps}Growth times and tangential period lengths of the fastest growing modes for the PS model with
cubic initial background field profile \eqref{quart},
initial penetration density $\rhoinit=10^{13} \gcmc$. 
Solid, diamonds: $\Bdfuenf$; dash-dotted, squares: $\Bdzwei$; 
dashed, triangles: $\Bdeins$. For further explanations see \figref{tiperqfp}.}
\end{center}
\end{figure}
\begin{figure}[h]

\vspace{-1.4cm}
\begin{center}
\epsfig{file=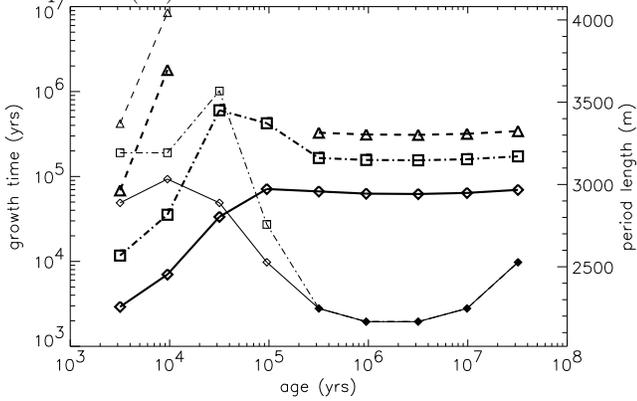, width=\linewidth}
\caption{\label{tiperops}As before, but with
 heptic initial background field profile \eqref{oct}.}
\end{center}
\end{figure}
%
%\enlargethispage{\baselineskip}
\begin{figure}[h]

\vspace{-1.6cm}
\begin{center}
\epsfig{file=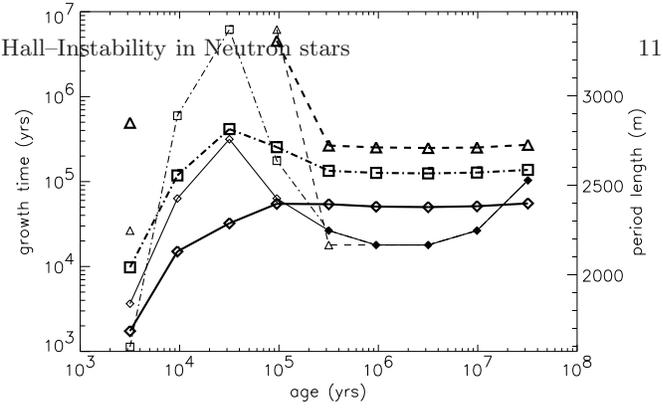, width=\linewidth}
\caption{\label{tipercps}As before, but with sinusoidal initial background field profile \eqref{cos}.}
\end{center}
\end{figure}
%
%\pagebreak
\begin{table*}[t]
\begin{center}
\begin{tabular}{clrrrrrr}   			
		\hline
		\hline\\*[-3mm]
$\Bd$ (G)\\*[1mm]
		\hline\\*[-2mm]
 &profile type             &  \multicolumn{2}{c}{cubic}       &  \multicolumn{2}{c}{heptic}     &  \multicolumn{2}{c}{sinusoidal} \\*[1mm]
		\cline{2-8}\\*[-2mm]
 &$\rhoinit (\gcmc)$ & \multicolumn{1}{c}{$10^{13}$} & \multicolumn{1}{c}{$10^{12}$} &\multicolumn{1}{c}{$10^{13}$} &\multicolumn{1}{c}{$10^{12}$} &\multicolumn{1}{c}{$10^{13}$} &\multicolumn{1}{c}{$10^{12}$}\\*[1mm]
		\hline\\*[-2mm]
\multirow{3}{8mm}{\hspace{2mm}\mbox{\begin{sideways}$\quad\quad5\times10^{13}$\end{sideways}}}
&global minimum   	& 850 	 &1,700	    &510     & 750    &550    &630	\\
&global maximum   	&256,060 &1,150,350 &107,650 &561,050 &53,590 &382,800	\\
%occurs at age			 &\multicolumn{6}{c}{----- $3\times 10^6$ -----}\\
&value at highest age	&84,880	 &326,520   &38,780  &185,630 &20,240 &122,370	\\*[1mm]	
		\hline\\*[-2mm]
\multirow{4}{8mm}{\hspace{2mm}\mbox{\begin{sideways}$\;10^{13}$\end{sideways}}}
&global minimum   		& 7,880	  		&74,900    		&3,470  		&8,290			&2,870			&5,620			\\
&global maximum   		&1,280,780		&12,767,800		&536,040		&3,711,550 		&266,740		&192,520		\\
&occurs at age			& $3\times 10^6$	&$10^6$   	&$3\times 10^6$ 	&$3\times 10^5$ 	&$3\times 10^6$	&$3\times 10^6$	\\
&value at highest age		&358,870  		&1,466,830 		&177,270		&720,110		&88,050			&517,920		\\*[1mm]
%gap				&	---		&$9.51\times 10^3$ --- $3.17\times 10^5$     	&	 ---    	&	---		&   ---	&  ---		\\	
		\hline
\end{tabular}
\end{center}

\vspace{2mm}
\caption{\label{compFP1} Growth times $\taugrow$ (yrs) of the FP model. For $\Bdfuenf$
the global minimum and maximum occur always at the ages of $3\times 10^3$ and $3\times 10^6$ yrs, resp.
For $\Bdeins$
the global minimum occurs always at the age of $3\times 10^3$ yrs. For the cubic profile and
$\rhoinit=10^{12}\gcmc$ no unstable modes exist between the ages $10^4$ and $3\times 10^5$ yrs.}
\end{table*}
\begin{table}[t]
\begin{tabular}{@{}c@{}l@{\hspace{2.5mm}}r@{\hspace{3mm}}r@{\hspace{3mm}}r@{\hspace{1mm}}}   			%p{.35\linewidth}p{.35\linewidth}}
		\hline
		\hline\\*[-3mm]
 $\Bd$ \\
 (G) \\*[1mm]
		\hline\\*[-1mm]
 &profile type             &  \multicolumn{1}{c}{cubic}      &  \multicolumn{1}{c}{heptic}     &  \multicolumn{1}{c}{sinusoidal} \\*[1mm]
		\hline\\*[-2mm]
\multirow{4}{7mm}{\hspace{1.5mm}\mbox{\begin{sideways}$5\times10^{13}$\end{sideways}}}
&global min.  		&3,500	 &2,920		&1,740 		\\
&global max.  		&227,490 &71,230	&55,330		\\
&occurs at age		&$10^5$	 &$10^5$	&$3\times 10^7$ \\
&val. at max. age	&135,300 &69,630	&55,330 	\\*[1mm]	
		\hline\\*[-2mm]
\multirow{6}{7mm}{\hspace{1.5mm}\mbox{\begin{sideways}$10^{13}$\end{sideways}}}
&global min.   		&590,900		&68,590			&248,650 		\\
&occurs at age     	& $3\times 10^6$	& $3\times 10^3$	&$3\times 10^6$ 	\\
&global max.   		&3,071,190		&1,796,610		&4,540,120		\\
&occurs at age		& $3\times 10^3$	& $10^4$		& $10^5$		\\
&val. at max. age	&658,290		&342,250		&271,080 		\\	
&gap			&$10^4\!\!\ldots10^5$	&$3\!\times\! 10^4\!\!\ldots10^5$&$10^4\!\!\ldots3\!\times\! 10^4$ \\*[1mm]
		\hline
\end{tabular}

\vspace{6mm}
\caption{\label{compPS1} Growth times $\taugrow$ (yrs) of the PS model. For $\Bdfuenf$
the global minimum occurs always at the age of $3\times 10^3$ yrs.}

\vspace{-4mm}
\end{table}
\noindent
and clearly smaller than the ones
for the cubic profile. However, the late stages reveal the sinusoidal profile to be the favorite; in the case
of $\rhoinit=10^{13} \gcmc$ it yields the smallest late stage growth time at all (20,240 yrs).
As a rule of thumb, the growth times for the cubic, heptic and  
sinusoidal profiles behave like 4:2:1 at late stages.
These ratios reflect even more or less quantitatively the relationship of the corresponding curvature parameters at the late stages, see \figref{FPcprofs}. 

\subsubsection{PS model}
A clearly pronounced and characteristic common property of the data shown in \figstoref{tiperqps}{tipercps}, is the
independence of $\taugrow$ on age for stages older than $3\times 10^5$ yrs. Further on, for these ages the dependence
of $\taugrow$ on $\Bd$ is {\em linear} to a high accuracy.
%(e.g., the ratio of the growth times for
%$\Bdeins$ and $\Bdfuenf$ lies in the interval $4.9\ldots 5.1$).
We explain the first effect by referring to the weak dependence
of the curvature parameter $(\alpha f')'/\eta$  
on age %in this range of ages 
(see \figsandref{PSqprofs}{PScprofs}) and the second by stating the dominance of the Hall terms in 
\eqref{indeqlin} in this range of ages.

In contrast, for the earlier stages the sensitivity of the growth times with respect to $\Bd$
is extraordinarily high. It is highest for the cubic profile for which at the age of $3\times 10^3$ yrs a scaling of $\Bd$ by 5 reduces the growth time
by almost 3 orders of magnitude. In general, with falling $\Bd$ a maximum of the growth time is forming between the ages of $10^4$ and $10^5$yrs.
%The ratio of the values for the cubic and the heptic profiles, is again $\approx 2$.
%which even breaks up to infinity for $\Bdeins$.
``Gaps" in age without any unstable modes are listed in Table \ref{compPS1}.

The background dynamics permissibility condition is not well satisfied at all the young stages, except
for $\Bdfuenf$. As the background field  and diffusivity profiles don't change significantly for
ages larger than $10^6$ yrs, the growth times are reliable from this age on.
A comparison of the different profile types with respect to the later stages yields the sinusoidal
one to be the most favorable for the instability, with the heptic one being close on its heels. %[(1:1.25)]. 

\subsubsection{Comparison FP --- PS model}  
From a final view on the results for both models we summarize the main differences 
between them as shown in \tablref{modscomp}.
\begin{table}[t]
\begin{tabular}{@{}p{.325\linewidth}@{\hspace{4mm}}p{.28\linewidth}@{\hspace{1mm}}p{.35\linewidth}@{}}
			 	    \hline
			 	    \hline\\*[-2mm]
			 	&   \multicolumn{1}{c}{FP}  	&  \multicolumn{1}{c}{PS}\\*[1mm]		  
			 	    \hline\\*[-2mm]
\mbox{dependence on age} 	 
during rapid cooling 	 	&  falling			&  constant	  	\\*[4mm]
			 	    \hline\\*[-2.5mm]
\mbox{age at which $\taugrow$}
is maximum (yrs) 		& some $10^{6}$ 		&  some $10^5$ 		\\*[4mm]
			 	    \hline\\*[-2.5mm]
\mbox{minimum $\taugrow$} in late stages (yrs)& $\approx$ 20,000& $\approx$ 55,000      \\*[4mm]
			 	    \hline\\*[-2.5mm]
\mbox{sensitivity of $\taugrow$}
with respect to $\Bd$         & \mbox{linear to}\linebreak
			        \mbox{weakly nonlinear}		& \mbox{extremely nonlinear} \linebreak
			                                          \mbox{at early ages,} \linebreak
							          \mbox{linear from $10^5$yrs on} \\*[1mm]

			 	    \hline\\*[1mm]
\end{tabular}
\caption{\label{modscomp} Comparison of the results for $\taugrow$ of the FP and the PS models}

\vspace{-3mm}
\end{table}

As a striking {\em common} feature of a majority of all the specific models one finds the very short growth
times at the youngest ages.
This is in contradiction to our assumption, introduced in \citet{GR02} and based on the magnitude of
$\ombt$, that the Hall
instability switches on only after a considerable cooling down of the NS, i.e.,
only after some $10^4$ to some $10^5$ yrs. XXX An early onset of
the instability could have a noticeable impact on the distribution of the magnetic field strengths across the NS population.
Namely, if we had a growth time of, say, 500 yrs at the age of 3,000 yrs 
changing to, say, 1,000 yrs at the age of 10,000 yrs one would expect an
essential part of the background field energy to be consumed by the unstable modes during some 10,000 yrs
after birth. XXX 
However, we see at least two reasons to 
suggest a very careful use of these ``early" growth times. First, during the early stages the instability
may benefit from perhaps unphysically high values of the curvature parameter inherited from our initial background fields.
For this, the value of the curvature parameter reached even infinity at $\rho=\rhoinit$ because the background field profile $f(z)$ shows a kink there.
For the later stages, instead, we are optimistic that this heritage lost its
significance due to the long lasting period of smoothing diffusion.
Second, even if the value of the curvature parameter could be regarded as realistic,
one had to take account of the high sensitivity of the conductivity with respect to
temperature during the hot young stage of the NS. Likewise, the high sensitivity of $\taugrow$ with 
\begin{figure*}[t]
\begin{center}
\begin{tabular}{@{\hspace{-.1cm}}c@{\hspace{.8cm}}c}
\epsfig{file=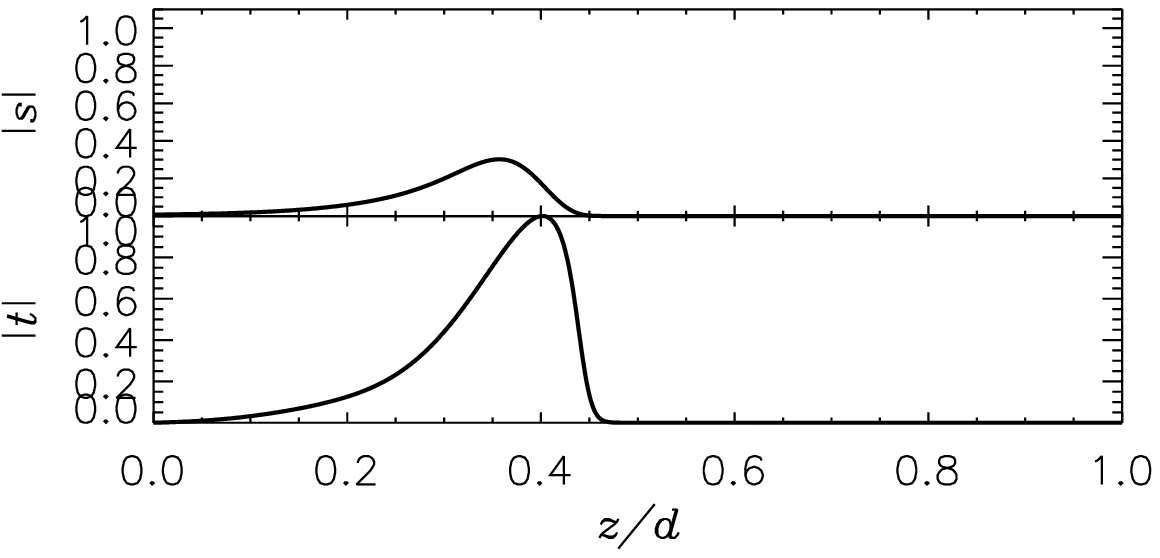, width=.48\linewidth} &
\epsfig{file=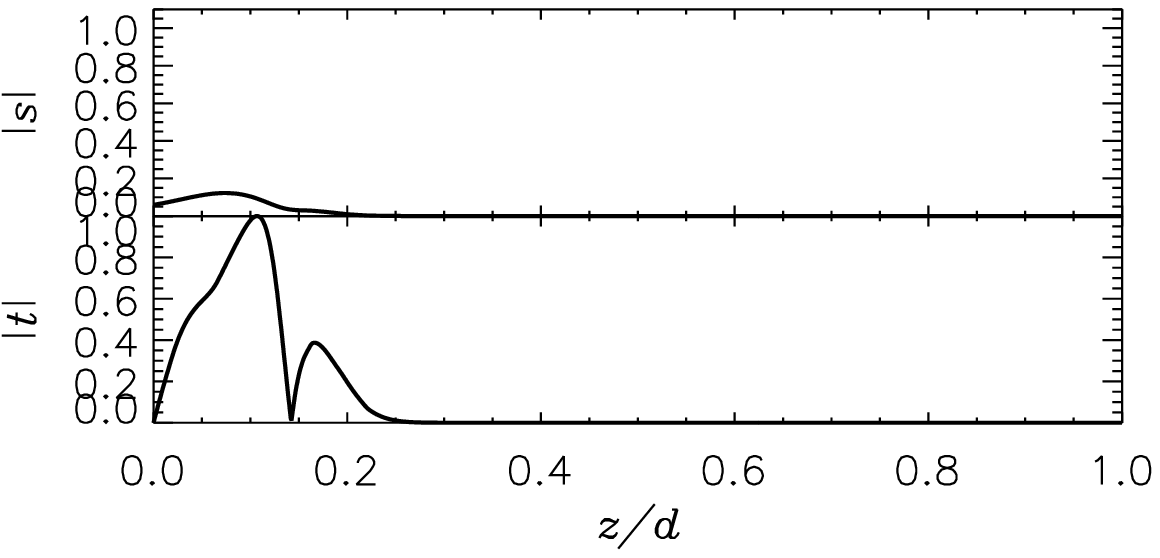, width=.48\linewidth}\\*[-1mm] 
\epsfig{file=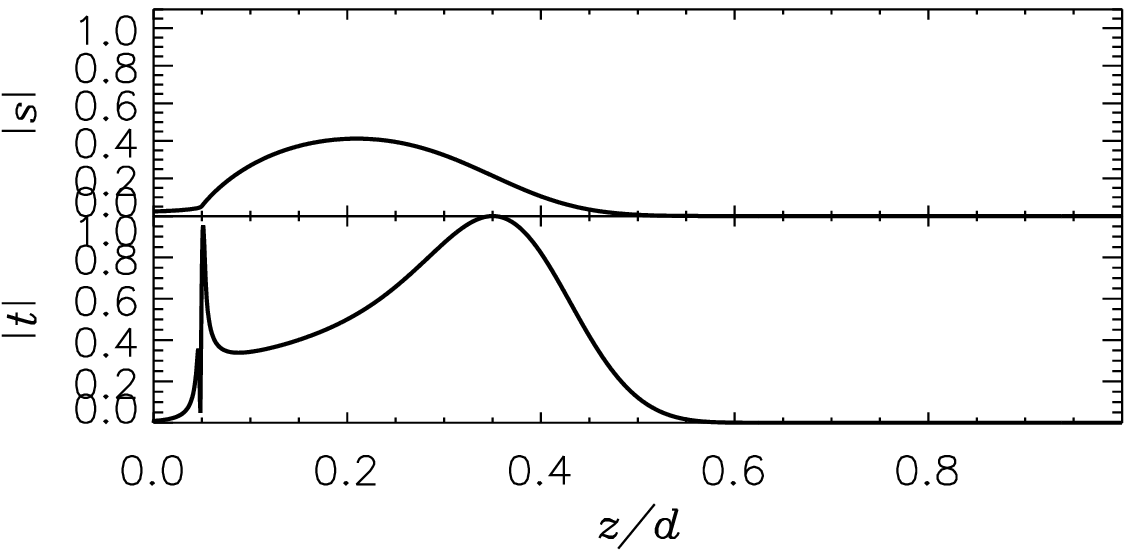, width=.48\linewidth} &
\epsfig{file=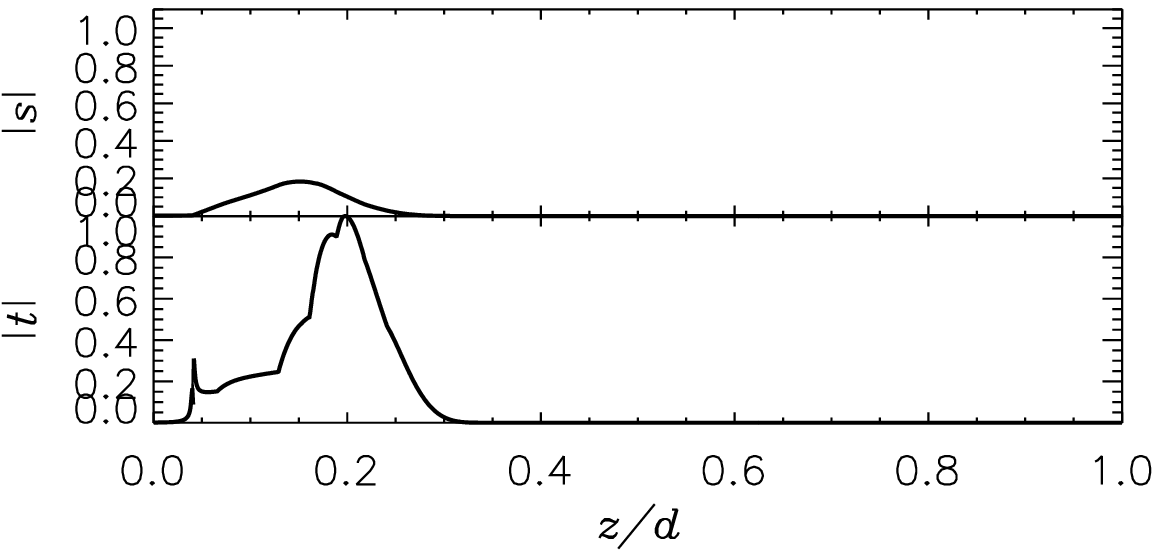, width=.48\linewidth} 
\end{tabular}
\caption{\label{eigfunc} Moduli of the scalars $s$ and $t$ of
the fastest growing mode for the FP (left) and PS (right) model with the sinusoidal
initial background field profile \eqref{cos}, initial field $\Bdfuenf$,
initial penetration density $\rhoinit=10^{13} \gcmc$ at the ages of $3\times 10^3$ yrs (above) and $3\times 10^7$ yrs
(below).}
\end{center}
\end{figure*}
respect to $\Bd$ and, hence, to the curvature parameter depending on conductivity, should be considered.
Thus, due to the enhanced local heating by the ``hot spots" of the unstable modes (see \figref{joule})
an efficient self--limitation of the mode amplitudes or even a switching--off of the instability
may take place. Perhaps, this happens far before a significant part of the
background field's energy is transferred to unstable modes.

\subsection{Eigenfunctions}
Figure~\ref{eigfunc} shows some typical eigenfunctions of the poloidal and toroidal scalars,
$s$ and $t$, for the most favorable background field profiles, that is, the sinusoidal 
ones with $\rhoinit=10^{13} \gcmc$, $\Bdfuenf$ at the earliest and latest stages, respectively.
Table \ref{compeigfunc} lists major qualitative differences of the eigenfunctions for the
two NS models.

In general all eigenfunctions show one pronounced (primary) maximum in both $|s|$ and $|t|$ 
with two additional (secondary) local maxima of $|t|$ close to the vacuum boundary for 
the two (FP) and five (PS) last ages, respectively.
The global maxima of $|s|$ lie a bit closer to the vacuum boundary than the corresponding
ones of $|t|$.
\begin{table*}[t]
\begin{tabular}{p{.2\linewidth}p{.36\linewidth}p{.36\linewidth}} %fuer 2spaltige Version
%\begin{tabular}{p{.2\linewidth}p{.48\linewidth}p{.48\linewidth}}  %fuer Referee-Version
\hline
\hline\\*[-2mm]
                 & \multicolumn{1}{c}{FP model} 	&    \multicolumn{1}{c}{PS model}	\\*[1mm]
\hline\\*[-2mm]
primary maximum \hfill\linebreak 
of $|s|$ and $|t|$ & \mbox{almost fixed near $z/d=0.4$ except at last age}
                   \mbox{when closer to surface:}\linebreak
                   \mbox{at $z/d=0.2$ for $|s|$ and $z/d=0.35$ for $|t|$}		&    almost fixed near $z/d=0.2$\\*[9mm]
			 	    \hline\\*[-2.5mm]
secondary maxima 
of $|t|$         & \mbox{lower than corresponding primary maximum}
                   \mbox{except at last age;}			   & \mbox{always lower than primary maximum}\linebreak\linebreak 		
                     			                             ``early" secondary maximum at
                                                                     $z/d=0.1$ at age $3\times 10^3$ yrs				\\*[2mm]
                   
			 	    \hline\\*[-2.5mm]
inner zero of $|t|$& \mbox{emerges at age $3\times 10^6$ yrs, moves from}
                      \linebreak
                      $z/d=0.2$ to $z/d=0.05$ with growing age 	&    \mbox{emerges at age $3\times 10^5$ yrs, moves from}
                                                                     \linebreak
                                                                     $z/d=0.07$ to $z/d=0.04$ with growing age;\linebreak
                                                                     \mbox{``early" zero at $z/d=0.14$ at age $3\times 10^3$ yrs.}\\*[1mm]                                      
\hline\\*[3mm]
\end{tabular}

\caption{\label{compeigfunc}Comparison of qualitative features of the eigenfunctions
(cf. \figsref{eigfunc}{eigfunc2}).} 
\end{table*}

With one exception, the secondary maxima are lower than the primary ones. They are sharpening with
growing age and $\Bd$, and emerge
immediately below and above a zero of $t$ which exists from the age of $3\times 10^5$ yrs on and
moves towards the vacuum boundary with aging.
In order to discuss this feature, we stress at first
that because of its clear convergent behavior with respect to
grid size it has to be considered as a physical fact rather than a numerical artifact. 
Since the zero of $t$ mimics a vacuum boundary (see \eqref{BCst}), it is intriguing to relate
the peak beneath it to the similar peak right beneath the vacuum boundary in the homogeneous density
model \citep[see][]{RG02}. 

Obviously, the perturbation fields concentrate 
mainly in regions of high values
of $\Bzero$ and/or the curvature parameter, respectively. 

To judge whether the field structures of the eigenmodes might be considered small--scale, we state
first that their major radial scale corresponds with the major radial scale of the corresponding background
field. The tangential scales can simply be identified with (half of the) tangential period lengths $\lambda$, which again
reflect the radial background field scale.
When comparing with the crust thickness, there are surely no small scales, except
the structures around the secondary maxima. But, when comparing with the NS perimeter it is well 
justified to claim small--scaleness since for the FP model the $\lamax$ lie
between 700 and 1000 m, whereas the NS's perimeter is 67~km. For the PS model we have  $\lamax=1,500 \ldots 4,500$ m
with a perimeter of 103~km.    

Because the non--smoothness of the conductivity profiles in the PS model
is not directly expressing itself in the background field profiles, one could expect the same for the eigenfunctions.
Indeed, the eigenfunctions of the first four ages don't show such signatures, but those of the five
last ages do. 
This fact is only partly explainable by the sharper kinks in the $\eta$ profiles of the last four
ages in comparison with those of the first five (see \figref{diffprofs}). It strengthens again the impression
that there is a qualitative difference between `earlier' and `later' stages of the PS model.

\begin{figure*}[t]
\begin{center}
\begin{tabular}{@{\hspace{-.7cm}}c@{\hspace{-.5cm}}c@{\hspace{-.2cm}}c}
\epsfig{file=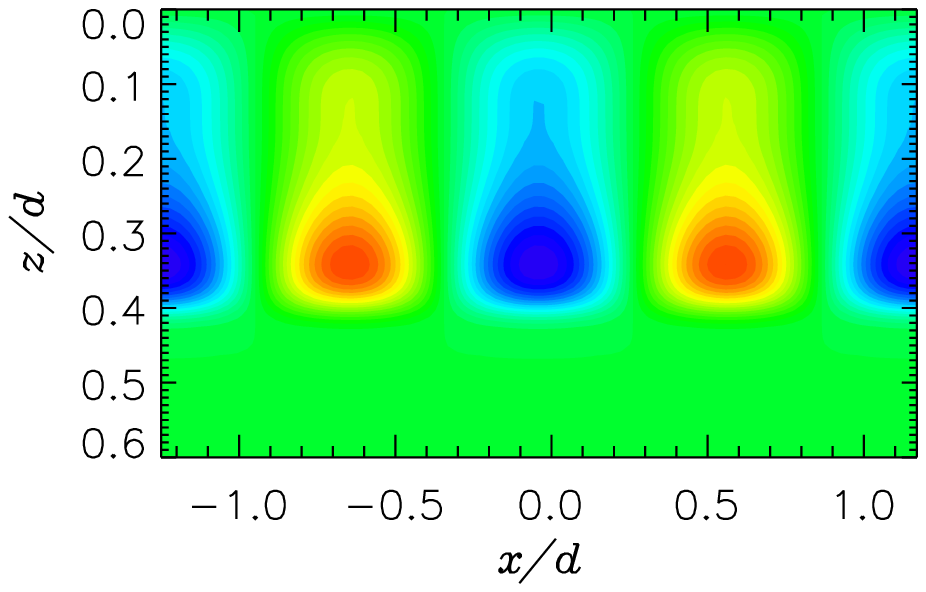 , width=.535\linewidth}\hspace{-.535\linewidth}%
\epsfig{file=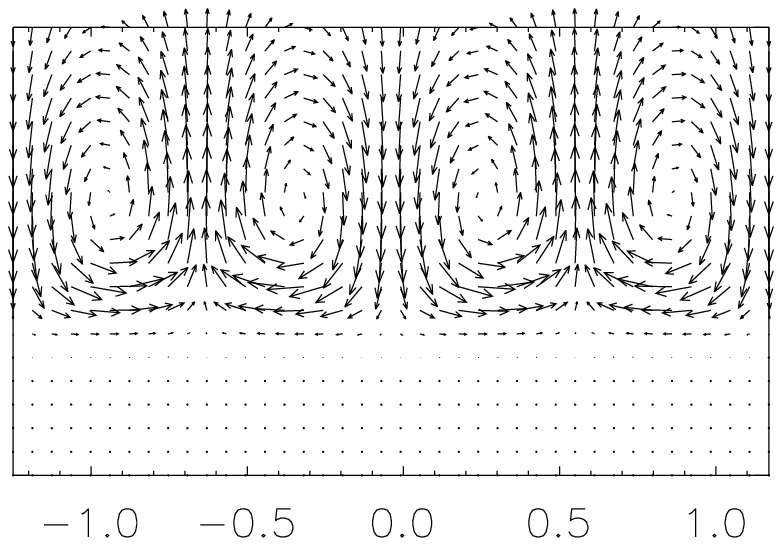, width=.535\linewidth} &  
\raisebox{.162\linewidth}{\large$3\times 10^3$} &
\epsfig{file=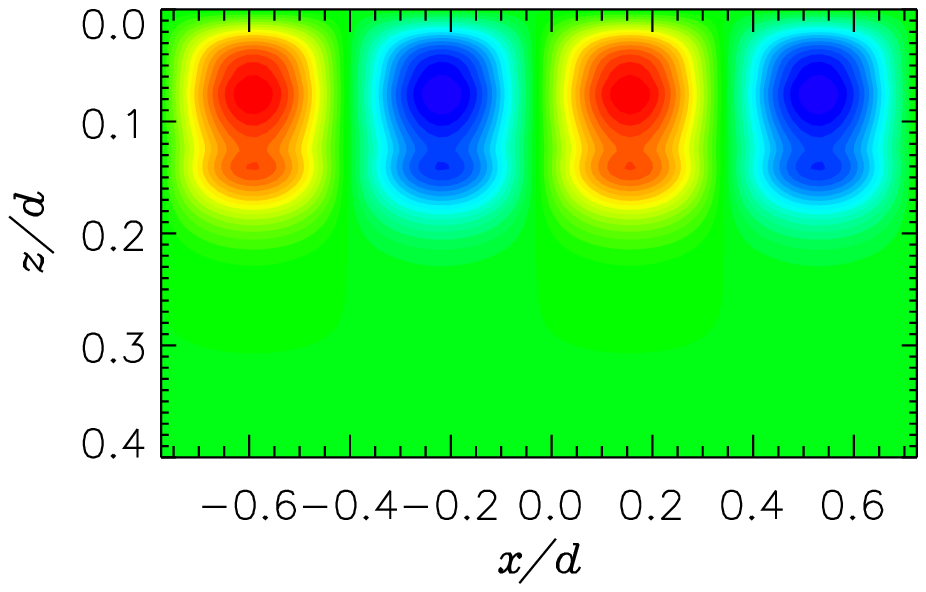 , width=.535\linewidth}\hspace{-.535\linewidth}%
\epsfig{file=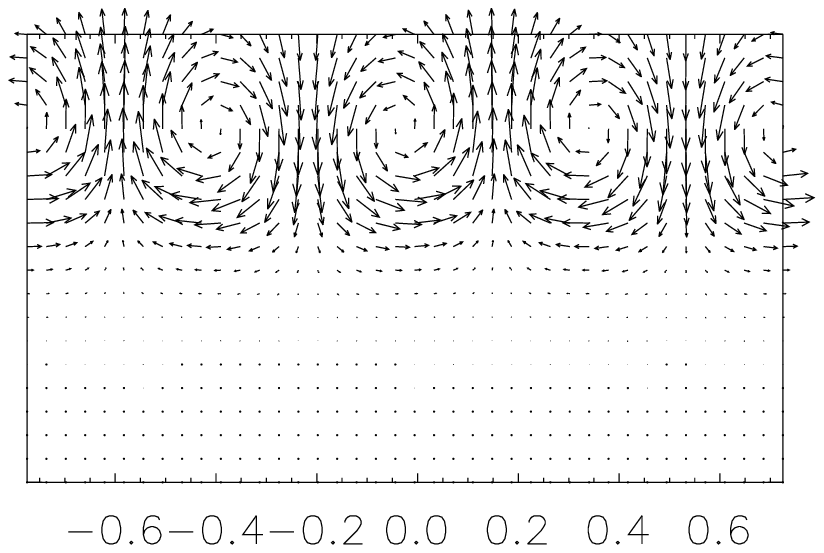, width=.535\linewidth}\\*[-1mm]
\epsfig{file=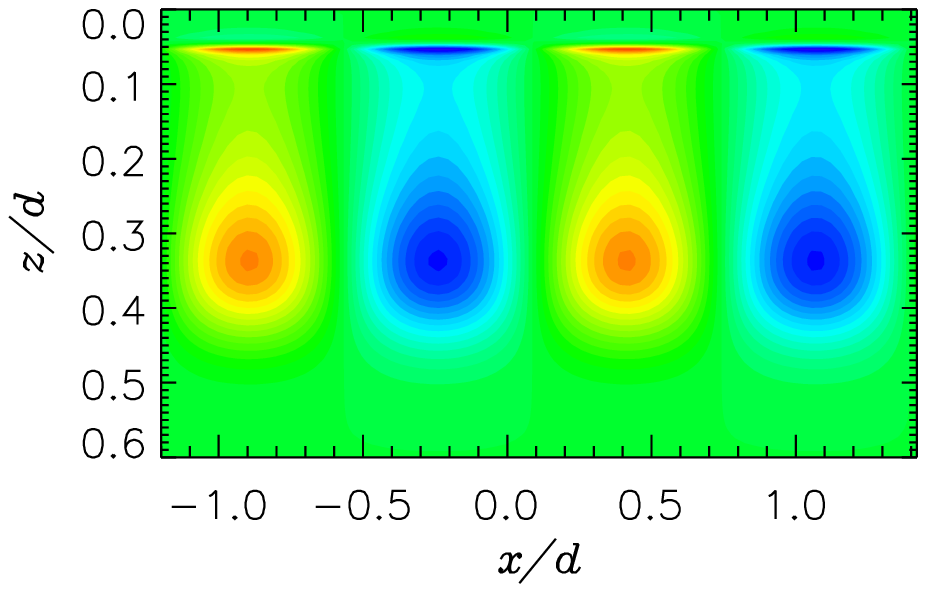 , width=.535\linewidth}\hspace{-.535\linewidth}%
\epsfig{file=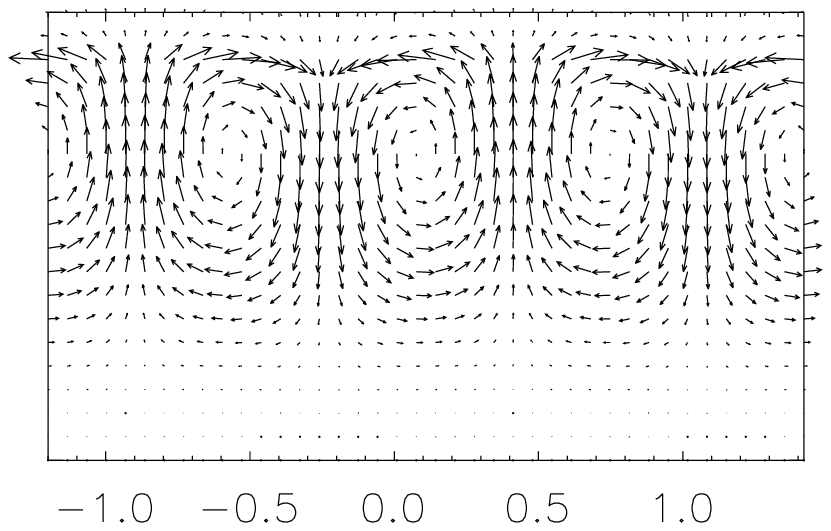, width=.535\linewidth} &  
\raisebox{.162\linewidth}{\large$3\times 10^7$} & 
\epsfig{file=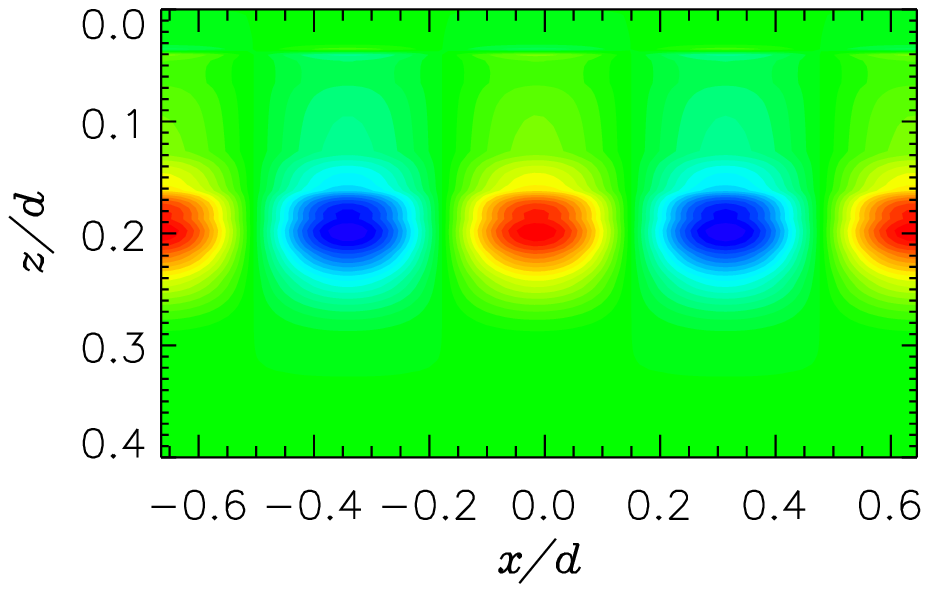 , width=.535\linewidth}\hspace{-.535\linewidth}%
\epsfig{file=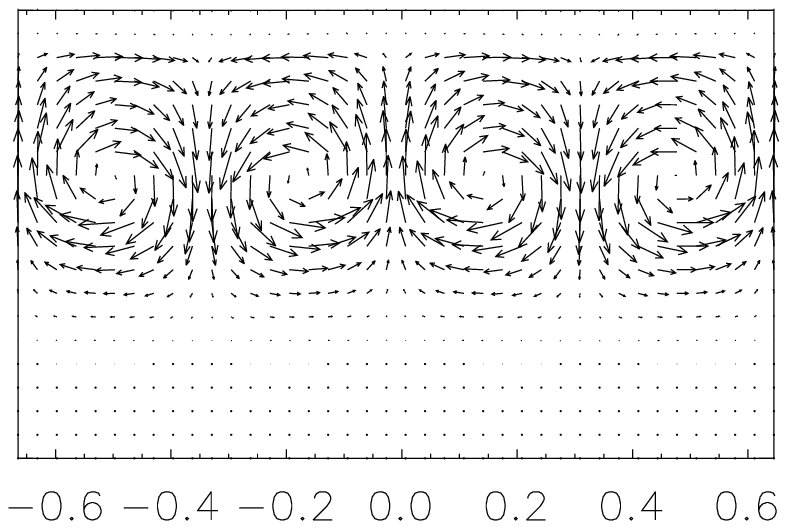, width=.535\linewidth}\\*[-1mm]
\epsfig{file=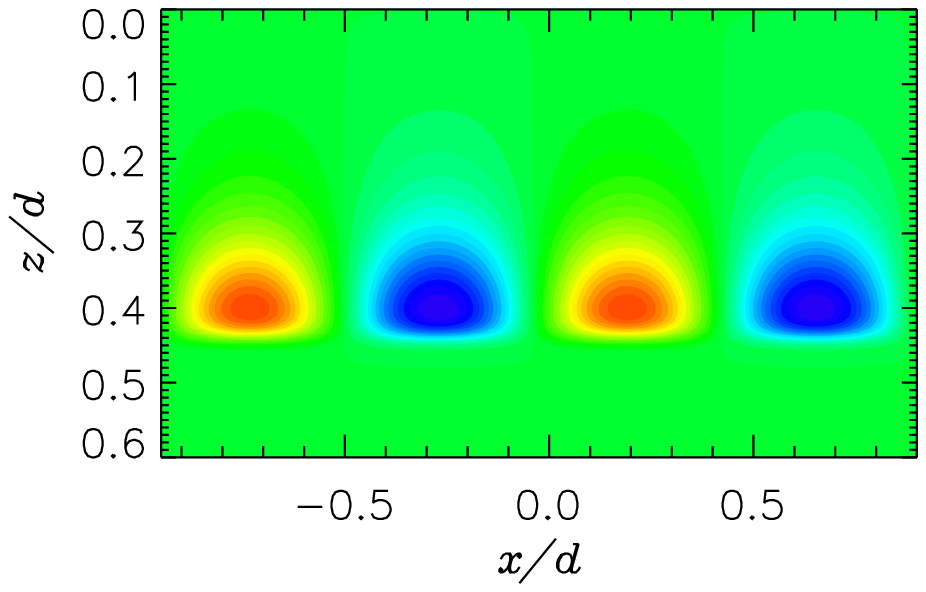 , width=.535\linewidth}\hspace{-.535\linewidth}%
\epsfig{file=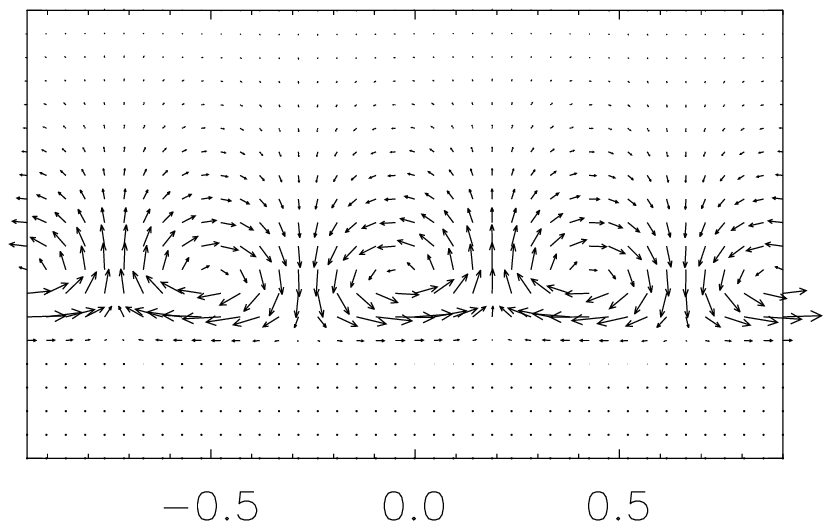, width=.535\linewidth} &
\raisebox{.162\linewidth}{\large$3\times 10^3$} &
\epsfig{file=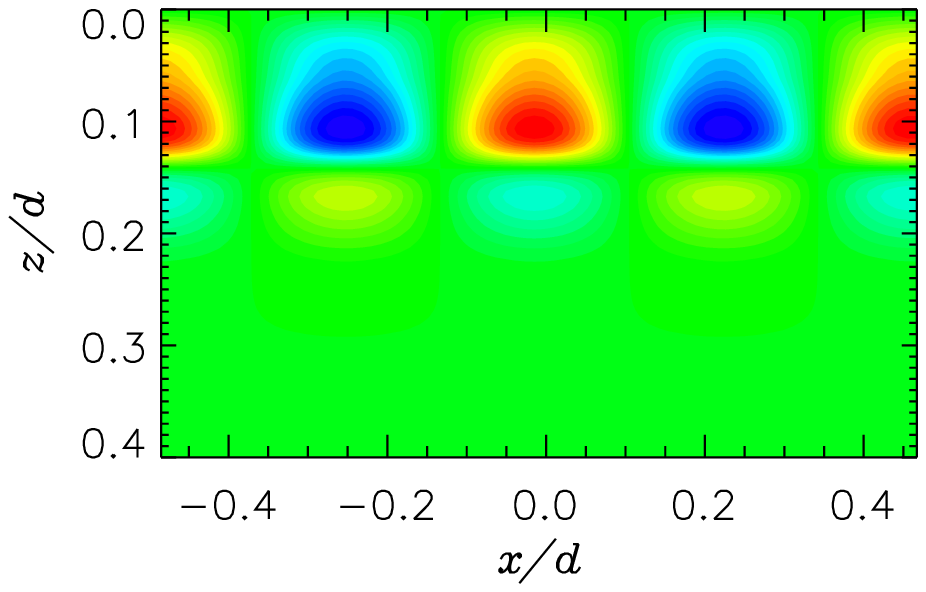 , width=.535\linewidth}\hspace{-.535\linewidth}%
\epsfig{file=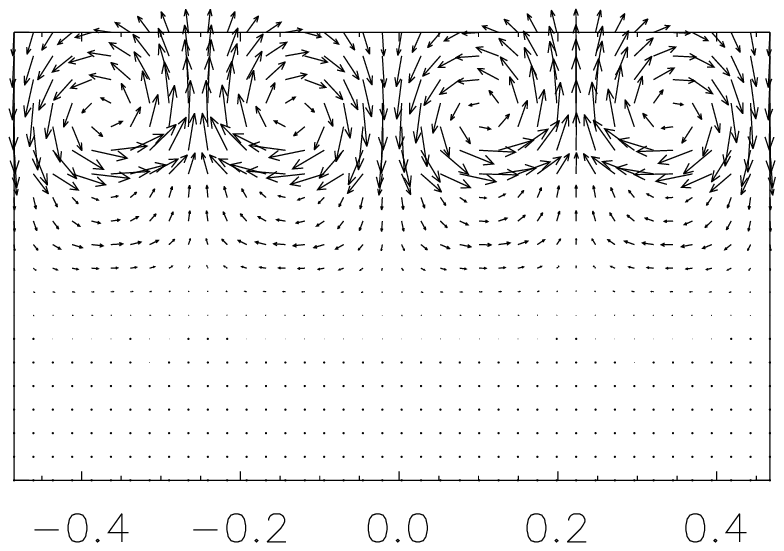, width=.535\linewidth}\\*[-1mm]
\epsfig{file=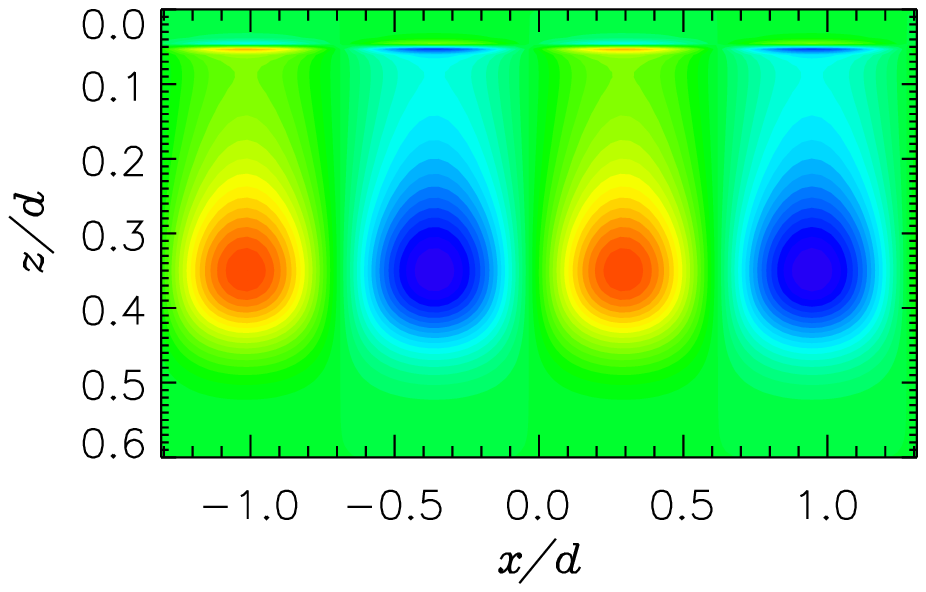 , width=.535\linewidth}\hspace{-.535\linewidth}%
\epsfig{file=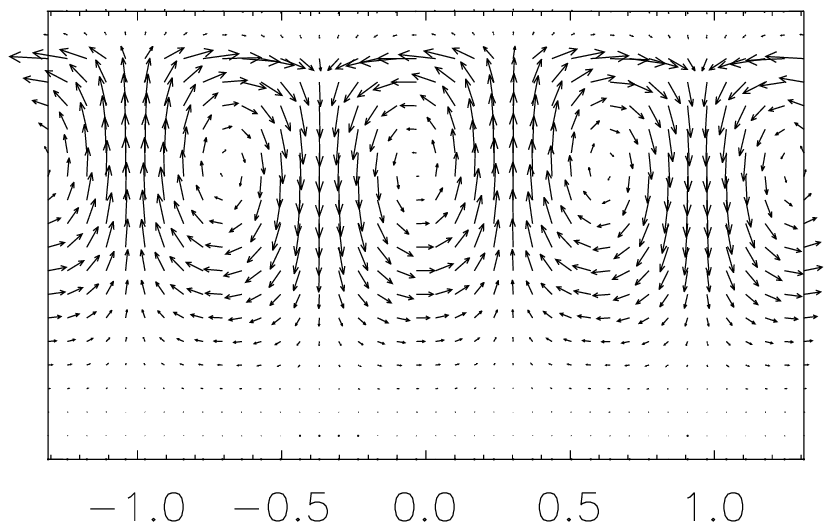, width=.535\linewidth} &
\raisebox{.162\linewidth}{\large$3\times 10^7$} &
\epsfig{file=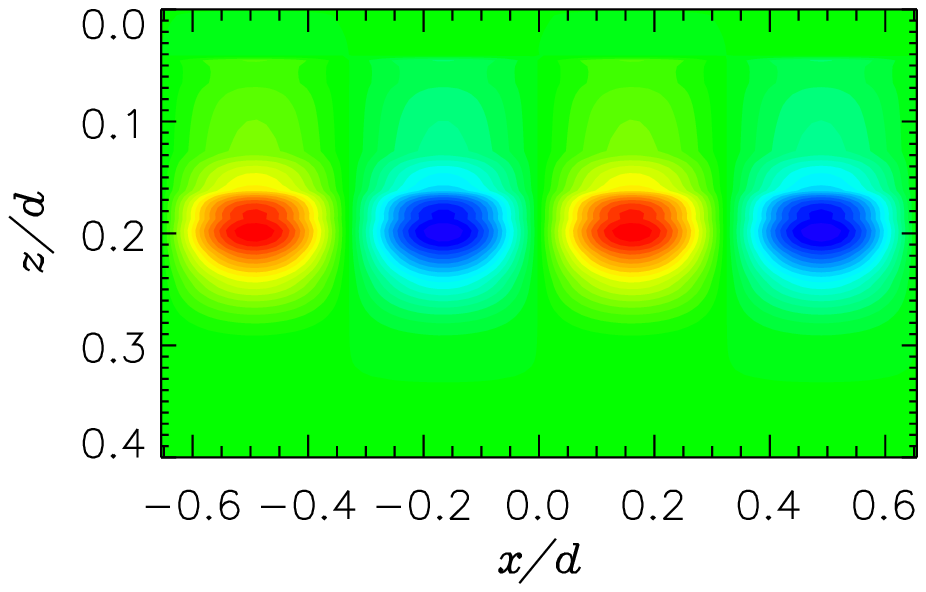 , width=.535\linewidth}\hspace{-.535\linewidth}%
\epsfig{file=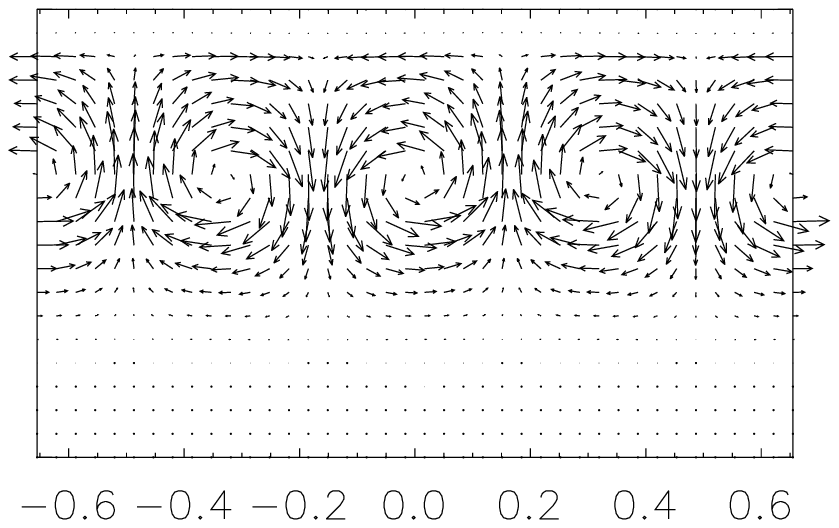, width=.535\linewidth}
\end{tabular}
\caption{\label{eigfunc2} Fields of
the fastest growing modes for the FP (left) and PS (right) model,
initial field $\Bdfuenf$,
initial penetration density $\rhoinit=10^{13} \gcmc$.
Upper two rows: cubic \eqref{quart},
lower two rows: sinusoidal \eqref{cos} initial background field profile.
Ages as indicated in the middle (in yrs). The field component perpendicular
to the paper plane is indicated via color encoding.}
\end{center}
\end{figure*}

Figure~\ref{eigfunc2} shows perturbation field structures for selected cases. As a general tendency
the fields of the FP model reach deeper layers of the crust and are less concentrated in $z$--direction than those of the PS model.
At early stages the fields tend to be relatively stronger outside the slab in comparison
with their maxima inside. Note the very sharp maxima of $b_y$ near the surface
for the latest stages of the FP model, which are connected with the occurrence 
of secondary maxima in the toroidal scalar $t$ (see \figref{eigfunc}).    

\section{Summary and Conclusions}
\label{conclus}
The results presented here prove that the instability, originally found in a slab of
uniform density,
does persist when a typical stratification of density and
chemical composition as present in NS crusts, and a realistic background field are employed.

We found that independent of the specific NS--model,
the strength and structure of
the background field, the growth times are
smallest at early stages of the
NS, reaching then a maximum at a model depending intermediate
age and become
smaller again in the process of further cooling.
In some cases (e.g., PS model, $\Bdeins$) the 
instability disappears for too low background
field strengths in a period of intermediate ages.

In order to understand this behavior, note that during the early period the crustal crystal
is so hot that a large number
of phonons is excited. Hence, the electron
relaxation time $\tau_e$ is
relatively small, causing in turn a small
magnetization parameter $\ombt$.
But, competing with that, the magnetic field, initially confined to a shallow
crustal layer had not yet time enough neither to decay
nor to diffuse deeper into
the crust therefore still preserving the curvature of its initial profile. 
It seems, that this effect dominates
the relative smallness of $\ombt$. 
In turn, during late, cool stages 
$\tau_e$ is much larger but the curvature of the
background field profile is due to its progressing diffusion and
decay considerably smaller. (See \figstoref{FPcprofs}{PScprofs}).

Quite general, our conclusion in \citet{RG02} that the appearance of the instability
depends {\it both} on a sufficiently strong
curvature of the background field profile
{\it and} a sufficient large magnetization
parameter $\ombt$ has thus been confirmed. However, it turned out that
in models with stratification one has to include
the derivative of the Hall--coefficient into the notion of `curvature' by introducing
a properly defined curvature parameter (see Sect.~2). It estimates the ratio of the most relevant
terms in the governing equations \eqref{indeqlin}.
 
The differences between growth times {\em within} the set of early stages (instants labelled 1 to 5 in \figstoref{FPcprofs}{PScprofs})
or late stages (instants labelled 7 to 9), respectively, are well reflected by the curvature parameter. 
But, unfortunately, it fails when comparing growth times between the two sets as it predicts the shortest
growth rates for the late stages.
In both NS models studied, there seem to exist 
subtle and hidden features of the coefficients and background field profiles
which cause a qualitative contrast between early and late stages not understood
up to now.  

Our results give rise to the idea that the
Hall--instability will act relatively early in the NS's
life. Hence, it could
reduce and smooth out the background
field in its progress, so that later on (say after $10^5$ yrs) the conditions for the
occurrence of the instability are no longer given.

On the other hand we found (mainly for the PS model) a high sensitivity of the `early' growth times with respect to the
background field strength. An increase by a factor of~$5$ can
cause a decrease of the
growth time by almost three orders of magnitude.
Recalling the importance of the profile curvature, we conclude that details of the background field (strength and structure) 
determined by the generation process of the NS magnetic
field may play a crucial role with respect to appearance and vigour of the instability at early stages.
Since these processes are far from being completely understood up to now, the only detail one can discuss
is the initial radial extent of the generated fields. 
Generation
by a thermoelectric instability in the relatively thin liquid shell forming the later
crust surely provides radial profiles with a lot of curvature. But whether they are really instability--friendly
has to be put in question as a small initial penetration depth of the background field ($\rhoinit=10^{12}\gcmc$) turned
out to be non--favorable. 
Considering alternatively typical structures of convection in a nascent NS \citep[see, e.g.,][]{KJM96,FH00}
one might conclude, that the assumption of bigger initial penetration depths 
could be appropriate for magnetic fields generated by a proto--NS dynamo. 
In all, we suggest to make use of the `early' growth times with great care only.
 
The late stages, in contrast, are
increasingly less influenced by peculiarities of the somewhat arbitrary assumptions
on the initial fields. At these stages the
dependence of the growth times on the background field strength is undramatic (close to linear).
So we think that we can rely 
safely on the growth times obtained at ages $\gtrsim10^5$ yrs.      

If the Hall--instability sets on at late stages, small--scale perturbations amplified 
at the expense of the background field may survive for a long time. To estimate
it, let us first look onto the Ohmic decay times for typical perturbations.
Their scale lengths are in the order of $5\times
10^4\ldots10^5$ cm, and their field maintaining currents
circulate in a depth of $2\dots5\times 10^4$ cm
corresponding to a density of about $10^{12}$~g cm$^{-3}$.
Taking into account that at an age of
$10^5\ldots10^6$ years the conductivity in that 
region is in the order of $10^{26}\ldots10^{27}$
s$^{-1}$, we arrive at values of $10^7...10^8$ years.
XXX However, these perturbations together with the background
field are subject to the Hall effect. Following a usual argumentation, one 
would therefore expect that the real decay times are shorter than the ohmic ones.
This conclusion would be based on the idea, that the Hall effect, although being conservative on its
own,  leads to an accelerated decay of a large scale field because energy is
flowing along the Hall cascade towards small scales the dissipation rate of
which is higher than that of the large scale field. However, this is not the whole truth
for a situation in which energy is already concentrated in a small scale.
Along with a Hall cascade to even smaller scales dissipating more quickly another one to larger scales 
can occur which dissipate
more slowly. Whether or not such a double-sided cascade leads to an overall
acceleration of the decay of the original small-scale mode must remain open. XXX 

\begin{figure}[t]
\epsfig{file=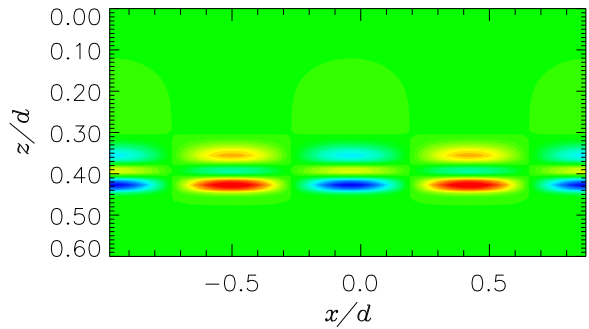, width=\linewidth}
\caption{\label{joule} Perturbation of the Joule heat sources density 
 $(\eta/2\pi)\curl\Bvec_0\cdot\curl\bvec$ in arbitrary units corresponding to the 
 perturbation field of the left column, third row of \figref{eigfunc2}. Green to red --- positive,
 green to blue --- negative deviations from the background heat sources.}
\end{figure}

The possible occurrence of relatively persistent small--scale field structures at
the NS surface together with the (episodically) accelerated decay of the large--scale background field
caused by the energy transfer to small--scale modes during their rapid growth 
represent the most important aspects of the Hall--instability 
with respect to observations. 
Possible observable consequences are therefore
\begin{itemize}
\item {\em deceleration} of the NS spin--down due to the {\em accelerated} magnetic field
      decay. XXX Note, that the real extent and duration of the latter can only be determined
  by analyzing the nonlinear stage of the instability including saturation. XXX
\item a hotter NS surface as a consequence of enhanced Joule heating due to the concentration of magnetic energy in the small
scales of unstable eigenmodes
\item glitches and bursts as a consequence of enhanced Lorentz forces making crust cracks
more probable
\item emergence of radio subpulses due to the existence of small--scale
structures in the vicinity of the magnetic pole.
\end{itemize}
A discussion of the first three effects can be found in \cite{GR02}. We will not repeat
it here since all the general arguments hold true for the Hall--instability in a stratified
crust, too.
However, some modification seems to be necessary with respect to the enhancement of Joule heating,
since in our present results the case of ``hot spots"
close to the surface is not the prevailing one  (\figref{joule}). Thus, the argument, that small--scale
temperature features due to localized Joule heating will be washed out and remain
unobservable in the light curve seems to be the more valid. On the other hand, 
an observable rise in the average surface temperature in comparison with the standard cooling
due to the additional ``deep" Joule heating is likely at least if the instability
acts during the phase for which the standard model predicts a rapid cooling ($10^5\ldots10^6$ yrs).

A necessary ingredient of the pulsar vacuum gap model of \citet{RS75} is the existence of
relatively strong and small--scaled
poloidal magnetic field structures just above the surface of the polar cap.
In a forthcoming paper we will demonstrate that the characteristics of such a
field as derived by \citet{GMG03} can be found with some of the unstable modes
presented here.

\begin{acknowledgements}
M.R. and U.G. gratefully acknowledge financial support by the Arbeitsamt Berlin
and the hospitality of the AIP. D.K. is thankful to the Humboldt foundation for the fellowship
realized with the hospitality of the AIP and to the support of grant
Nr. 03-02-17522 of RFBR.
\end{acknowledgements}

\bibliographystyle{aa}

\end{document}